\documentclass[prd,twocolumn,superscriptaddress,floatfix,nopacs,preprintnumbers,nofootinbib]{revtex4}
\usepackage[utf8]{inputenc}
\usepackage[T1]{fontenc}

\usepackage{graphicx}
\usepackage{float}
\usepackage[normalem]{ulem}
\usepackage{xcolor}
\usepackage{adjustbox}
\usepackage{mathrsfs}
\usepackage{amssymb,bm}
\usepackage{amsmath}
\usepackage{mathtools}
\usepackage{physics}
\usepackage{slashed}
\usepackage{verbatim}
\usepackage{ragged2e}
\usepackage{caption}
\usepackage{subcaption}
\usepackage{multirow}
\usepackage{array}
\usepackage{tikz}
\usepackage{lipsum} 
\usepackage{soul}
\newcommand{\nc}{N_\mathrm{c}}
\newcommand{\aem}{\alpha_\mathrm{em}}

\newcommand{\bt}{\mathbf{b}}
\newcommand{\qso}{Q_{\mathrm{s},0}} 
 
\newcommand{\rt}{\mathbf{r}}

\newcommand{\kt}{\mathbf{k}}

\newcommand{\pt}{\mathbf{p}}

\newcommand{\xbj}{x_\mathrm{Bj}}

\newcommand{\lqcd}{\Lambda_\text{QCD}}

\newcommand{\mve}{MV$^e$ }

\newcommand{\as}{\alpha_\mathrm{s}}

\usepackage[breaklinks,colorlinks,citecolor=citcolor,urlcolor=blue,linkcolor=lcolor]{hyperref}
\definecolor{lcolor}{rgb}{0.5,0,0}
\definecolor{citcolor}{rgb}{0,0.3,0.0}
\definecolor{teal}{rgb}{0.0, 0.5, 0.5}

\begin{document}

\title{Probing gluon saturation through inclusive hadron production in DIS}

\author{Carlisle Casuga}
\email{carlisle.doc.casuga@jyu.fi}
\affiliation{
Department of Physics, University of Jyväskylä,  P.O. Box 35, 40014 University of Jyväskylä, Finland
}
\affiliation{
Helsinki Institute of Physics, P.O. Box 64, 00014 University of Helsinki, Finland
}
\author{Swaleha Mulani}
\email{swaleha.n.mulani@jyu.fi}
\affiliation{
Department of Physics, University of Jyväskylä,  P.O. Box 35, 40014 University of Jyväskylä, Finland
}
\affiliation{
Helsinki Institute of Physics, P.O. Box 64, 00014 University of Helsinki, Finland
}
\author{Heikki Mäntysaari}
\email{heikki.mantysaari@jyu.fi}
\affiliation{
Department of Physics, University of Jyväskylä,  P.O. Box 35, 40014 University of Jyväskylä, Finland
}
\affiliation{
Helsinki Institute of Physics, P.O. Box 64, 00014 University of Helsinki, Finland
}

\begin{abstract}
We investigate  gluon saturation effects in semi-inclusive deep inelastic scattering (SIDIS) at small-$x$ within the Color Glass Condensate framework. We compute the  SIDIS cross section at leading order in the dipole picture, expressing it in terms of the dipole-target scattering amplitude. 
We validate our framework by comparing with the charged hadron spectra measured at HERA. 
Our predictions for the EIC indicate that saturation effects result in significant nuclear suppression in the SIDIS cross section, and 
can provide  complementary  constraints on the initial condition for the Balitsky-Kovchegov evolution extracted from inclusive DIS data.
These results demonstrate that SIDIS measurements at the EIC are sensitive to nonlinear QCD dynamics in the saturation regime.

\end{abstract}

\maketitle
\section{Introduction}
\label{sec:intro}

Total cross section measurements in Deep Inelastic Scattering (DIS) processes at HERA have revealed that parton densities grow rapidly towards the small parton longitudinal momentum fraction $x$~\cite{H1:2015ubc}. Although this growth cannot continue indefinitely without violating unitarity and at some point non-linear phenomena (e.g. gluon recombination) have to tame this growth, so far no conclusive evidence for such saturation effects has been seen~\cite{Morreale:2021pnn}. 

One potential approach to search for signatures of gluon saturation  is to look for more differential observables. 
This is typically done within the Color Glass Condensate framework~\cite{Gelis:2010nm,Albacete:2014fwa,Blaizot:2016qgz} that provides a convenient description of the QCD dynamics in the high parton density (or high collision energy) domain. In particular, in this region the center-of-mass energy (or momentum fraction $x$) dependence of the cross section is perturbative, and given by the B-JIMWLK evolution equation~\cite{Balitsky:1995ub,Kovchegov:1999yj,Kovchegov:1999ua,Jalilian-Marian:1996mkd,Jalilian-Marian:1997qno,Jalilian-Marian:1997jhx,Jalilian-Marian:1997ubg,Kovner:2000pt,Weigert:2000gi,Iancu:2000hn,Iancu:2001ad,Ferreiro:2001qy}.
Phenomenological studies of gluon saturation phenomena in DIS  include, for example, diffractive cross sections~\cite{Kowalski:2008sa,Lappi:2023frf,Fucilla:2023mkl,Fucilla:2022wcg,Boussarie:2016ogo,Boussarie:2014lxa}, diffractive vector meson production~\cite{Mantysaari:2025ltq,Kovchegov:2023bvy,Kowalski:2006hc,Boussarie:2026qzy,Boussarie:2024pax,Boussarie:2024bdo,Boussarie:2016bkq,Besse:2013muy}, dijet production~\cite{Mantysaari:2019hkq,Altinoluk:2015dpi, Boussarie:2019ero}, and heavy meson production~\cite{Gimeno-Estivill:2025rbw,Cheung:2024qvw}. 

SIDIS provides a complementary approach to searches of gluon saturation phenomena~\cite{Goncalves:2019owz,Iancu:2020jch}. Compared to  DIS structure function measurements, in SIDIS the hadron transverse momentum and rapidity can be varied continuously in addition to Bjorken-$x$ and $Q^2$, providing potentially more stringent tests of the gluon saturation picture of the CGC. For example, we demonstrate in this work that future SIDIS data from electron-nucleus scattering can provide some additional constraints to the inference of the non-perturbative initial condition for the B-JIMWLK evolution equation.

Single inclusive hadron production in DIS has been measured at small-$x$ in electron-proton collisions at HERA~\cite{H1:1996muf, H1:2004xgw}. Low-energy SIDIS measurements have also been performed e.g. by the HERMES collaboration at DESY for ep~\cite{HERMES:2007plz,HERMES:2012uyd} and eA~\cite{HERMES:2011qjb} collisions, and by the CLAS collaboration at JLAB~\cite{CLAS:2021jhm}. In the next decade, accurate multi-dimensional measurements of the nuclear modification factor for SIDIS can be performed at the Electron-Ion Collider~\cite{Aschenauer:2019kzf,AbdulKhalek:2021gbh,Duan:2026jrk}. Thanks to the high center-of-mass energy and luminosity, such studies can be expected to cover a wide phase space in hadron transverse momentum $\pt$, photon virtuality $Q^2$ and the nuclear mass number $A$, enabling one to probe transition from the dilute regime to the dense, saturated regime of  QCD.

The SIDIS cross section in the dipole picture has been derived  at leading order in $\as$ in~\cite{Mueller:1999wm,Kovchegov:2015zha}
 and at next-to-leading order (NLO) in~\cite{Bergabo:2024ivx,Bergabo:2022zhe, Altinoluk:2025dwd}. Furthermore, cross section for single inclusive jet production  in DIS has been analytically derived at NLO accuracy within the CGC framework in~\cite{Caucal:2024cdq}.  
In this work, we provide first realistic leading-order estimates for the nuclear modification factor in SIDIS, caused by saturation effects. The non-perturbative input in our calculation comes from fits to precise proton structure function data~\cite{Casuga:2023dcf,Lappi:2013zma}. When the dipole-proton scattering amplitude is generalized to the dipole-nucleus case using the optical Glauber model following the prescription of Ref.~\cite{Lappi:2013zma}, predictions for the nuclear modification factor are parameter-free, except for the minimal dependency on the applied fragmentation function.

This article is organized as follows. In Section~\ref{sec:setup}, we present the SIDIS cross section in the dipole picture in the case of proton and nuclear targets.  In Section~\ref{sec:results}, numerical results of this study are discussed in detail. In particular, first in Sec.~\ref{subsec:HERA} we validate our setup against the HERA SIDIS data. Then, predictions for $\gamma^*+A$ scattering, to be measured at the EIC,  are discussed in detail in Section~\ref{subsec:gamma_A scattering results}. Conclusions of our study are presented in Sec.~\ref{sec:conclusions}. In Appendix~\ref{append:A},
we quantify the uncertainty in the SIDIS cross section originating from the uncertainty of the dipole-target scattering amplitude.

\section{Setup}
\label{sec:setup}
Inclusive quark production in photon-nucleus scattering, $\gamma^* + A \to q + X$, where $A$ denotes a nucleus with mass number $A$, can be written as a convolution of the photon wave function,  describing the $\gamma^*\to q\bar q$ splitting, and the dipole-target scattering amplitude $N$,  describing the eikonal interaction of the quark-antiquark pair with the target proton or nucleus. The cross section for semi-inclusive quark production in DIS is obtained by integrating over the phase space of the final-state antiquark\footnote{We work in the high-energy limit where the phase space of the second quark is unbounded, but note that an approach for going beyond this limit has been recently proposed in Ref.~\cite{Bertilsson:2026vtu}.}.  The cross section for antiquark production is identical.
This cross section has been derived, for example, in Refs.~\cite{Mueller:1999wm,Kovchegov:2015zha,Marquet:2009ca}. 
For a longitudinally polarized photon, it reads
%

\
\begin{multline}
\label{eq:xs_L}
\frac{\dd[2]\sigma^{\gamma^* + A \to q + X}_{\mathrm{L}}}{\dd[2]{\pt_q} \dd{z_q}} = \frac{\aem \nc  }{(2\pi)^4}   e_f^2 \int \dd[2]{\kt} \dd[2]{\bt} \tilde S(\kt,\bt,x_g)  \\
\times 8 \epsilon^2  z_q (1-z_q)  
 \bigg[ \frac{1}{\epsilon^2 + \pt_q^2} - \frac{1}{\epsilon^2 + (\pt_q - \kt)^2}\bigg]^2 . 
\end{multline}
Here, $\aem = 1/137$ is the electromagnetic fine structure constant and $\nc =3$ is the number of colors.\\ 
Similarly, for a transversely polarized photon the cross section is
\begin{multline}
\label{eq:xs_T}
    \frac{\dd[2]{\sigma_{\mathrm{T}}}^{\gamma^* + A \to q + X}}{\dd[2]{\pt_q} \dd{z_q}} = \frac{\aem \nc }{(2\pi)^4}   e_f^2 \int \dd[2]{\kt} \dd[2] {\bt}  \tilde S(\kt,\bt,x_g) \\ 
\times  \bigg\{ 2 \left( z_q^2 + (1-z_q)^2\right)\bigg[ \frac{\pt_q}{\epsilon^2 + \pt_q^2} - \frac{\pt_q -  \kt }{\epsilon^2 + (\pt_q - \kt)^2}\bigg]^2 \\
 + m_f^2
\bigg[ \frac{1}{\epsilon^2 + \pt_q^2} - \frac{1}{\epsilon^2 + (\pt_q - \kt)^2}\bigg]^2 \bigg\}. 
\end{multline}
Here $\pt_q$ is the quark transverse momentum and $z_q$ is the fraction of the large photon plus momentum carried by the quark (or antiquark) that fragments into a measured hadron. Furthermore, $\bt$ is the impact parameter between the target and the $q\bar q$ dipole, $e_f$ is the fractional charge of the quark, and $\epsilon^2=Q^2z_q(1-z_q)+m_f^2$, where $m_f$ is the quark mass, and $f$ refers to the quark flavor. 

The cross sections in Eqs.~\eqref{eq:xs_L} and \eqref{eq:xs_T} are written in terms of the momentum space dipole amplitude proportional to the dipole gluon distribution~\cite{Dominguez:2011wm}
\begin{equation}
\label{eq:S_k}
    \tilde S(\kt,\bt,x_g) =  \int \dd[2]{\rt} e^{i \kt \cdot \rt} \left[ 1-N(\rt,\bt,x_g) \right].
\end{equation}
The advantage of this momentum space formulation, similarly as e.g. in Ref.~\cite{Gimeno-Estivill:2025rbw}, is that the numerically challenging Fourier transform of the dipole amplitude can be computed separately using a numerical procedure described below. 

The longitudinal momentum fraction of the target gluons involved in the scattering is denoted by $x_g$. Following Ref.~\cite{Iancu:2020jch}, it can be approximated as  
\begin{equation}
\label{eq:xg}
    x_g = \xbj \left(1+ \frac{\pt_q^2}{Q^2 z_q} + \frac{\max \{Q^2z_q(1-z_q), Q_s^2\}}{Q^2(1-z_q)}   \right),
\end{equation}
where $\xbj$ is the Bjorken-$x$ and $Q_s^2$ is the saturation scale.
This corresponds to the fraction of the target longitudinal momentum needed to put the $q\bar q$ system on shell, with the transverse momentum of the unmeasured quark estimated as  $\max \{Q^2z_q(1-z_q), Q_s^2\}$. 
For the proton, we use the saturation scales extracted in Ref.~\cite{Casuga:2023dcf}. For nucleus, we use the estimated saturation scale at the center of the nucleus based on the optical Glauber model of Ref.~\cite{Lappi:2013zma}: $Q_s^2(\text{nucleus}) =  A T_A(0) \frac{\sigma_0}{2}Q_s^2(\text{proton})$. 
This approximation corresponds to the unobserved (anti)quark carrying a typical transverse momentum $\max \{Q^2z_q(1-z_q), Q_s^2\}z_q(1-z_q)$. Here $T_A$ is the transverse thickness function calculated from the Woods-Saxon distribution normalized such that $\int \dd[2]{\bt} T_A(\bt)=1$, and $\sigma_0/2$ is the proton transverse area extracted from the DIS fit~\cite{Casuga:2023dcf}.

The cross section for single inclusive hadron production is obtained by convoluting the partonic cross section with the fragmentation function. In this work we use the NNFF1.0 leading order fragmentation function~\cite{Bertone:2017tyb}  
and write the cross section as
\begin{multline}
\label{eq:ff_conv}
    \frac{\dd \sigma^{\gamma^* + A \to h + X}_{\mathrm{L,T}}}{\dd[2]\pt \dd{z_h}} = \sum_{f=q,\bar q} \int \frac{\dd{z_f}}{z_f^3} D_{f\to h}(z_f, \mu^2) \\ 
    \times \left.\frac{\dd[2]\sigma^{\gamma^* + A \to q + X}_{\mathrm{L,T}}}{\dd[2]{\pt_q} \dd{z_q}}\right|_{z_q=z_h/z_f, \pt_q = \pt/z_f }.  
\end{multline}
Here $z_h = z_f z_q$ is the fraction of the photon plus momentum carried by the produced hadron $h$, $z_f$ is the fraction of the quark momentum carried by the produced hadron, and $\pt$ is the hadron transverse momentum.
Unless otherwise specified, we calculate the sum of the longitudinal and transverse cross sections, where the partonic cross sections are given by Eqs.~\eqref{eq:xs_L} and~\eqref{eq:xs_T} respectively, and are convoluted with the fragmentation function according to Eq.~\eqref{eq:ff_conv}.
We evaluate the fragmentation function $D_{f\to h}$ at scale $\mu^2=\pt^2$, and the uncertainty related to this scale choice is quantified by varying the scale from $\mu^2=(\pt/2)^2$ to $\mu^2=(2\pt)^2$. As the measured hadron can originate from the quark or the antiquark, both possibilities are included in Eq.~\eqref{eq:ff_conv}, where sum runs over the light flavors $f = u,d,s$ and their corresponding antiquarks $\bar{u}, \bar{d}, \bar{s}$.

When comparing with the HERA data~\cite{H1:1996muf}, we compute hadron production at fixed rapidity $y$ in the photon-proton center-of-mass frame
\begin{multline}
    \frac{\dd \sigma^{\gamma^* + A \to h + X}_{\mathrm{L,T}}}{\dd[2]\pt \dd{y}} = \sum_{f=q,\bar q} \int \frac{\dd{z_f}}{z_f^3} D_{f\to h}(z_f, \mu^2) \\ 
    \times z_h\left.\frac{\dd[2]\sigma^{\gamma^* + A \to q + X}_{\mathrm{L,T}}}{\dd[2]{\pt_q} \dd{z_q}}\right|_{\pt_q = \pt/z_f },  
\end{multline}
where
\begin{equation}
    z_q = \frac{\sqrt{\pt^2 + m_h^2}}{z_f W}e^y.
\end{equation}
Here, $m_h$ is the hadron mass and $W$ is the center-of-mass energy of the photon-proton system. 

The remaining ingredient we need to specify in order to calculate the SIDIS cross section is the dipole-target scattering amplitude $N(\rt,\bt,x_g)$. The energy ($x_g$) dependence of the dipole is given by the Balitsky-Kovchegov evolution equation~\cite{Kovchegov:1999yj,Balitsky:1995ub}, but the initial condition for the evolution is non-perturbative. A typical approach is to parametrize the dipole at the initial $x_0=0.01$ using a McLerran-Venugopalan model~\cite{McLerran:1993ni} inspired parametrization, and infer its parameters from the HERA inclusive structure function data. We apply two different fits from Refs.~\cite{Lappi:2013zma,Casuga:2023dcf} where one first assumes that the impact parameter dependence in the dipole-proton scattering factorizes,
\begin{equation}
    \int \dd[2]{\bt} N(\rt,\bt,x) = \frac{\sigma_0}{2} N(\rt,x_g),
\end{equation}
and then parametrizes the impact-parameter independent initial condition for the dipole-proton amplitude   at $x_0=0.01$ as
\begin{equation}
\label{eq:dipole_param}
    N(\rt,x_g=x_0) = 1 - \exp\left[ -\frac{\rt^2 \qso^2}{4} \ln\left(\frac{1}{|\rt|\lqcd}+ e_c\cdot e\right) \right].
\end{equation}
Here, $\lqcd = 0.241$ GeV, $e$ is Euler's number, and $e_c$ is a dimensionless fit parameter. The term $e_c \cdot e$ regulates the logarithm at large dipole sizes. We note that the fragmentation functions are an independent input to our calculation, with their own $\as$ and flavor scheme~\cite{Bertone:2017tyb} that do not exactly match  the applied BK fits of Refs.~\cite{Lappi:2013zma,Casuga:2023dcf}. At the leading order accuracy of this work the resulting mismatch is a higher order effect, and it can be expected to mostly cancel in cross section ratios like the nuclear modification factor $R_\mathrm{pA}$. 

The dipole-proton scattering amplitude is generalized to the dipole-nucleus case by applying the optical Glauber model following Ref.~\cite{Lappi:2013zma}. This corresponds to solving the impact-parameter independent BK evolution equation independently at each $\bt$ with an initial condition
\begin{multline}
\label{eq:dipole_nucleus}
    N_A(\rt,\bt,x_g=x_0) = 1 - \exp \left[ -A T_A(\bt) \frac{\sigma_0}{2} \frac{\rt^2 \qso^2}{4} \right.  \\
    \times \left. \ln\left(\frac{1}{|\rt|\lqcd}+ e_c\cdot e\right)  \right].
\end{multline}
 By construction, this setup exhibits a trivial target nucleus mass number $A$ scaling of the cross sections in the dilute limit, i.e. no nuclear modification. As already mentioned above, $T_A(\bt)$ is the nuclear transverse thickness function obtained from the Woods-Saxon distribution normalized to $\int \dd[2]{\bt}=1$.
In the region where the saturation scale of the nucleus would fall below that of the proton, we use~\cite{Lappi:2013zma}
\begin{equation}
    N_A(\rt,\bt,x_g) = A T_A(\bt) \frac{\sigma_0}{2} N(\rt,x_g)
\end{equation}
which corresponds to no nuclear modification.

The $x$ dependence of the dipole amplitude is  obtained by solving the (impact-parameter independent) BK evolution from $x=x_0$ to $x=x_g$. As discussed above, in this work this evolution range is set by the typical invariant mass of the produced $q\bar q$ system. When deriving the cross sections~\eqref{eq:xs_L} and~\eqref{eq:xs_T}, one has integrated over the unobserved quark, and consequently this choice is not unique. The sensitivity of our results on this choice is quantified in Appendix.~\ref{append:evolutionscale}. We also note that the choice~\eqref{eq:xg} is not completely consistent with the applied DIS fit where the dipole amplitude is always evaluated at $x_g=\xbj$. In principle it would also be possible to use an evolution scale similar to \eqref{eq:xg} in DIS fits following Ref.~\cite{Bertilsson:2026vtu}, but no such fits have been published.

For $\xbj$ close to the initial condition of the BK evolution, a significant fraction of the cross section originates from $x_g>0.01$, see Eq.~\eqref{eq:xg}. In our numerical implementation, we freeze the dipole amplitude at its initial condition and set $N(\rt,\bt,x_g)=N(\rt,\bt,x_0=0.01)$ for $x_g>0.01$. Consequently, our numerical results close to $\xbj=0.01$ exhibit sensitivity to the prescription used to extrapolate the dipole amplitude beyond the fitted range. We demonstrate and quantify this uncertainty in Appendix.~\ref{append:evolutionscale}. 

As an initial condition, we use the ``MV'' model fit from Ref.~\cite{Lappi:2013zma} where one sets $e_c\equiv 1$ in~\eqref{eq:dipole_param} and considers $\qso^2$ and $\sigma_0/2$ as free parameters. 
In order to determine if current and future SIDIS measurements can provide complementary constraints to the BK fits, we also use the ``MV$^e$'' fit from Ref.~\cite{Casuga:2023dcf} where the infrared regulator $e_c$ is also taken to be a free parameter, which is found to improve the fit quality. Uncertainty estimates for the fit parameters are available for the MV$^e$ fit, and the corresponding uncertainty in the SIDIS cross section is quantified in Appendix~\ref{append:A}. Fit parameters used in our study are summarized in Table \ref{tab:ic_parameters}. Consistently with the applied fit setups, we also include only the three light quark flavors with $m_f=0.14\,\mathrm{GeV}$.

\begin{table}[tb]
\centering
\renewcommand{\arraystretch}{1.4}
\begin{tabular}{l c c c c c}
\hline\hline
\textbf{Model} & $Q_{s0}^2\ [\text{GeV}^2]$ &  $e_c$ & $\sigma_0/2\ [\text{mb}]$ & $C^2$ \\
\hline
MV  & $0.104$  & $1$ & $18.81$ & $14.5$ \\
MV$^e$ & $0.054$  & $58.532$ & $14.34$ & $5.198$ \\
\hline\hline
\end{tabular}
\caption{Initial condition parameters for the BK evolved dipole amplitudes used in this work. The MV and MV$^e$ model fits are from Refs.~\cite{Lappi:2013zma} and~\cite{Casuga:2023dcf} respectively. The parameter $C^2$ controls the coordinate space running coupling scale in the BK evolution and does not directly appear in the SIDIS calculation.}
\label{tab:ic_parameters}
\end{table}
The Fourier transformed dipole $\tilde S$ defined in Eq.~\eqref{eq:S_k} is computed as follows. At each $x$ (and $\bt$ when considering dipole-nucleus scattering), we take the numerical solution to the BK equation, and approximate it by a function    
\begin{multline}
\label{eq:dipolefitparam}
    N(\rt,\bt,x_g) = \Bigg\{ 1 - \exp \Biggl[ \Biggl(- \frac{(r^2 \qso^2)^\gamma}{4} \\
   \times  \ln\left(\frac{1}{r\lqcd}+ e_c \cdot e\right) \Biggr)^p \Biggr] \Bigg \}^{\frac{1}{p}}.
\end{multline}
The parameters $\qso^2$, $e_c$, and $p$ are determined separately at each $x_g$.
To ensure that both the small- and large-$r$ regions are weighted equally, we fit the logarithm of Eq.~\eqref{eq:dipolefitparam} to the logarithm of the numerical data.
We have verified that the resulting parametrization provides an accurate description of the dipole amplitude at all  evolution rapidities relevant to this work.  
Since the dipole amplitude has no angular dependence, the angular integral in Eq.~\eqref{eq:S_k} can be performed analytically, which gives
\begin{equation}
\label{eq:S_k_h}
    \tilde S(\kt,\bt,x_g) =  2\pi \int_{0}^{\infty} \dd{\rt} \rt J_{0}(|\kt|\rt)\left[ 1-N(\rt,\bt,x_g) \right].
\end{equation}

The remaining Hankel transform is evaluated using analytical parametrization given earlier and standard algorithms implemented in the \texttt{Hankel} package~\cite{murray2019hankel}.

\section{Results}
\label{sec:results}

\subsection{HERA baseline}
\label{subsec:HERA}

\begin{figure}[tb]
    \centering
    \includegraphics[width=.5\textwidth]{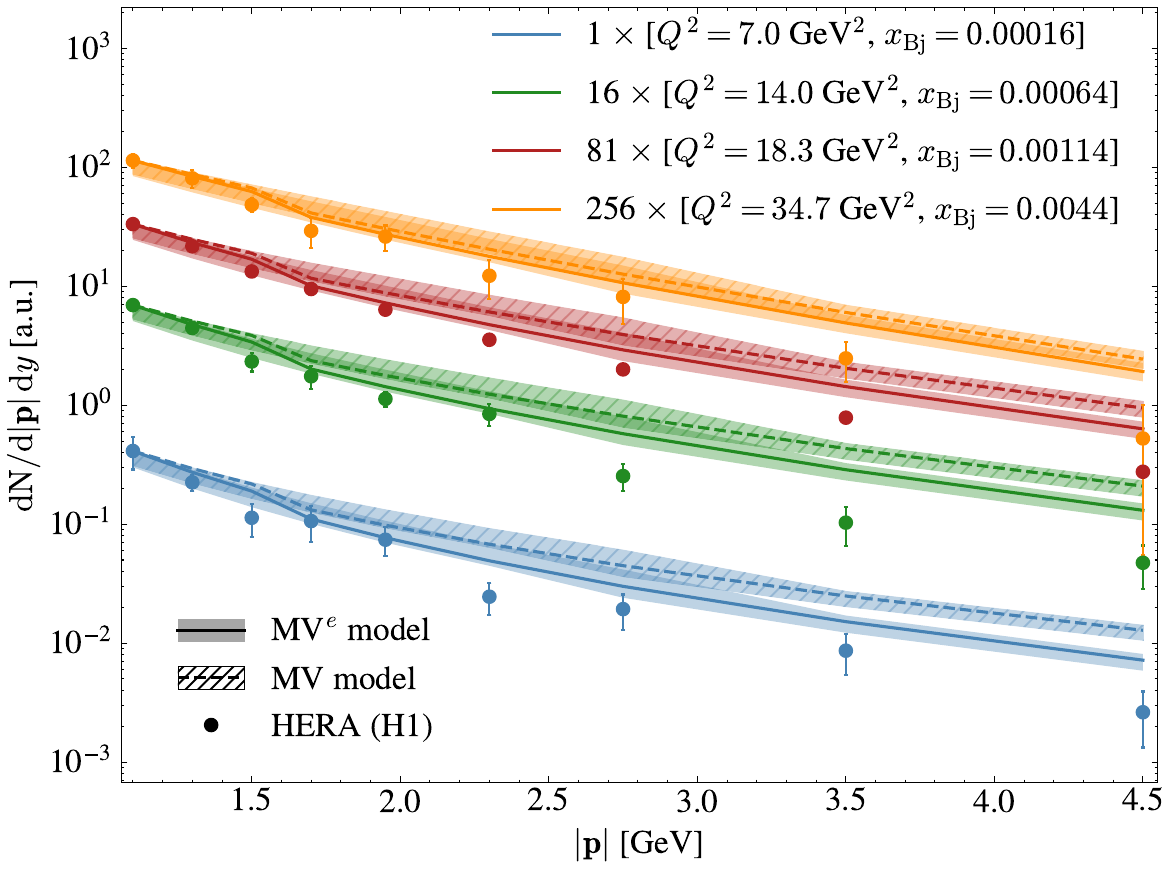}
    \caption{ 
    Charged hadron transverse momentum spectra (arbitrary units) in $\gamma^* + p \to h^\pm + X$ scattering in the photon-proton center-of-mass frame at $y=2$ compared with the HERA data~\cite{H1:1996muf}.    
    The uncertainty band corresponds to variation of the fragmentation function scale $\mu=0.5|\pt|\dots2|\pt|$. 
    Results are normalized to match the HERA data at lowest $|\pt|$. 
    For clarity, the data and calculations at the different ($Q^2, \xbj$) are multiplied by the factors indicated in the legend.
    }
    \label{plot:Plot_FF_Scale_Uncertainty_HERA_3Tables_gP_HERA_VS_MVe_VS_MV}
\end{figure}

We begin our analysis by comparing the charged
hadron transverse momentum spectra with the  H1 data~\cite{H1:1996muf}. We use both the MV and \mve fits for the dipole amplitude, and calculate the SIDIS spectrum in H1 kinematics in the small-$x$ region. The results are shown in Fig.~\ref{plot:Plot_FF_Scale_Uncertainty_HERA_3Tables_gP_HERA_VS_MVe_VS_MV}. 
As the differential HERA SIDIS data is not reported at the cross section level\footnote{We note that there is also single inclusive pion production in $e+p$ scattering data from HERA reported at the cross section level~\cite{H1:2004xgw}. However, that data is not presented  differentially in $\xbj, \pt$ and $y$ (or $z_h$) and as such comparison to that measurement is not presented here.  }, 
our results are normalized separately at each $Q^2$ bin to match the lowest-$|\pt|$ data points. 

The hadron spectra in DIS calculated using the MV and \mve dipoles are nearly identical in HERA kinematics at lower $|\pt|\lesssim 3.5\,\mathrm{GeV}$. Overall, a relatively good description of the HERA data is obtained, particularly in this lower $|\pt|$ region. However, at  higher transverse momenta $|\pt|\gtrsim 3.5\,\mathrm{GeV}$, the computed spectra are too hard. This is especially the case if the MV model dipole is used. Similar systematics has been previously observed in Ref.~\cite{Lappi:2013zma} when cross sections computed using similar BK initial condition parametrizations have been compared with particle spectra measured in proton-nucleus collisions.

The fragmentation function scale uncertainty, estimated by varying the scale $\mu$ by a factor of $2$, is comparable to the experimental uncertainties. For the \mve model, uncertainty estimates for the dipole amplitude are available and can be propagated to the  $\pt$ spectra. The resulting uncertainty is small, and we quantify it in more detail in Appendix~\ref{append:A}.
We conclude that the leading order dipole picture setup, constrained by the HERA structure function data, provides a reasonable description of the available small-$x$ SIDIS data and is therefore suitable for making realistic predictions for future EIC measurements.

\subsection{Predictions for $\gamma^*+A$ scattering}
\label{subsec:gamma_A scattering results}

\begin{figure*}
\centering

\begin{subfigure}[t]{0.49\textwidth}
\centering
\includegraphics[width=\linewidth]{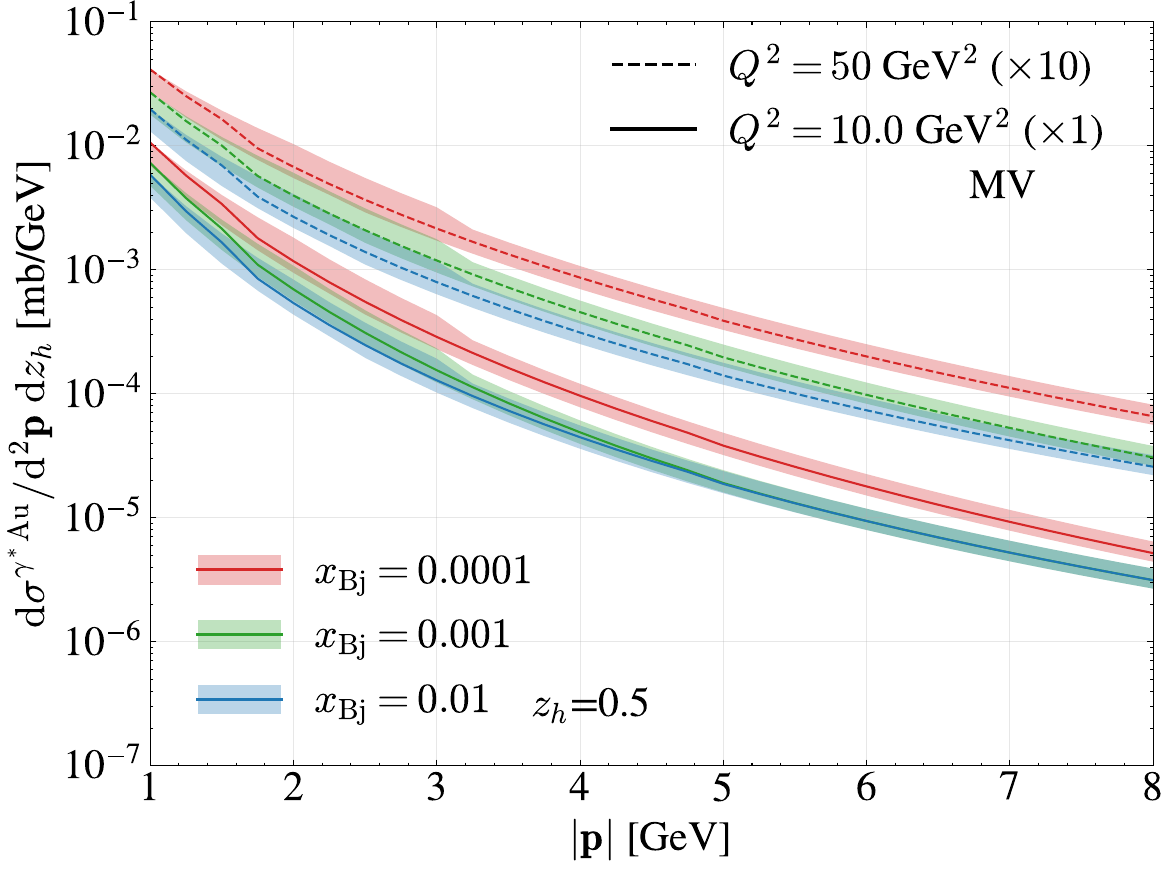}
\caption{MV}
\label{plot:Plot_xsec_A_FF_uncer_EIC_Q2_10_50_MV}
\end{subfigure}
\hfill
\begin{subfigure}[t]{0.49\textwidth}
\centering
\includegraphics[width=\linewidth]{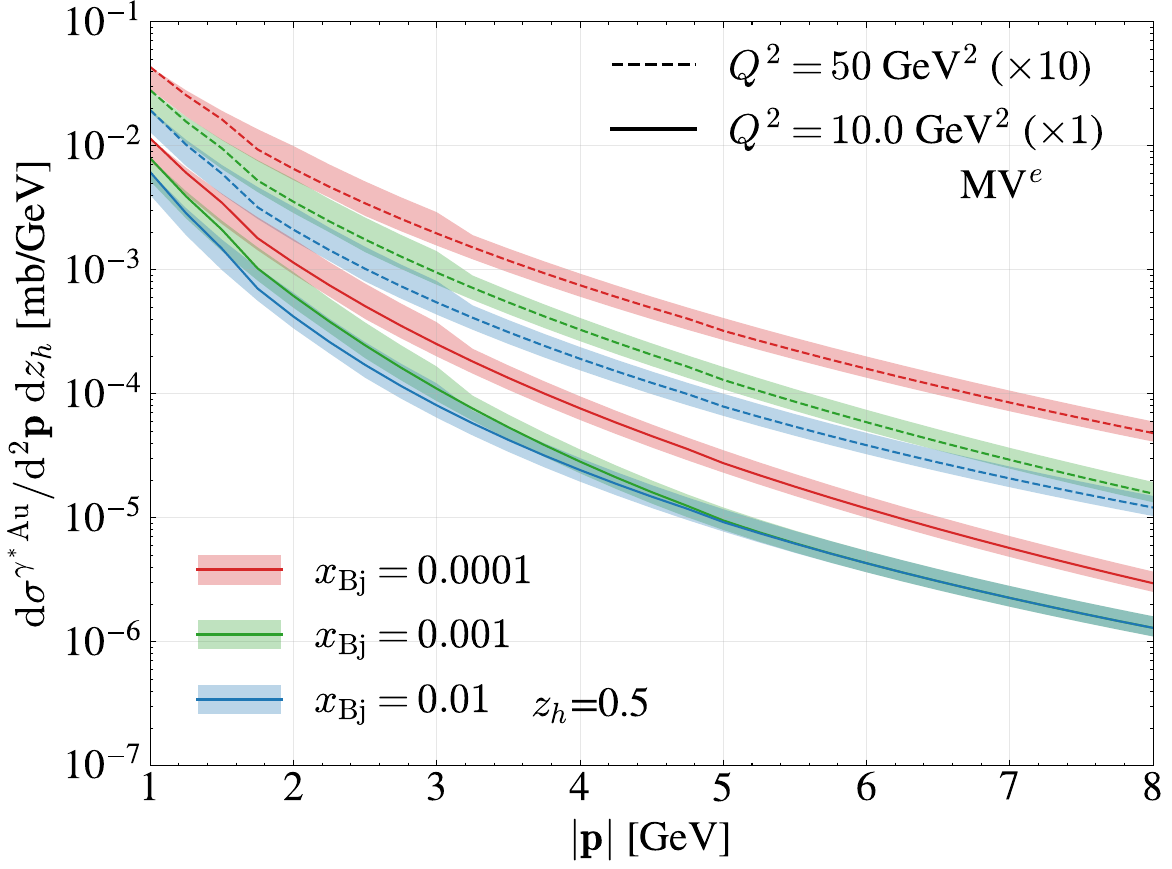}
\caption{MV$^e$}
\label{plot:Plot_xsec_A_FF_uncer_EIC_Q2_10_50_MVe}
\end{subfigure}

\caption{Differential cross section for $\gamma^{*} +\mathrm{Au}\to\pi^\pm + X$ scattering as a function of transverse momentum $|\pt|$, shown for selected $\xbj$ and $Q^2$ values at fixed $z_h = 0.5$. For clarity, the results at $Q^2$ = 50 GeV$^2$ are multiplied by 10.}

\label{fig:Plot_xsec_A_FF_uncer_EIC_Q2_10_100}
\end{figure*}
%
Next we study inclusive charged pion production in photon-nucleus scattering. The charged
pion production cross section as a function of pion transverse momentum $\pt$ in $\gamma^* +\mathrm{Au}$ scattering is shown in Fig.~\ref{fig:Plot_xsec_A_FF_uncer_EIC_Q2_10_100}. The results are calculated using both the MV (Fig.~\ref{plot:Plot_xsec_A_FF_uncer_EIC_Q2_10_50_MV}) and \mve (Fig.~\ref{plot:Plot_xsec_A_FF_uncer_EIC_Q2_10_50_MVe}) dipole amplitudes, and are found to be similar in both cases. 
The results are shown at fixed $z_h=0.5$, and we have confirmed that the shape of the spectra depends weakly on $z_h$. 

The relatively modest change in the cross section between $\xbj=0.01$ and $\xbj=0.001$ follows from the fact that, for both values of $\xbj$, a substantial fraction of the cross section originates from the region $x_g>0.01$,  where the dipole amplitude is extrapolated outside the fitted kinematical range and is taken to be independent of $x_g$. 
Consequently, the weak $\xbj$ dependence observed here close to $\xbj=0.01$ should be regarded as a numerical artifact rather than a genuine prediction of the physical $\xbj$-dependence, and suggests that our numerical results become less reliable close to $\xbj=0.01$. If a DIS structure function fit using an evolution variable consistent to Eq.~\eqref{eq:xg} were available and applied here with a consistent extrapolation to the $\xbj>0.01$ regime, this problem would likely be alleviated. The uncertainty related to the effective description of $x_g>0.01$ regime is quantified in Appendix~\ref{append:evolutionscale}.

\begin{figure*}[t]
\centering

\begin{subfigure}[t]{0.49\textwidth}
\centering
\includegraphics[width=\linewidth]{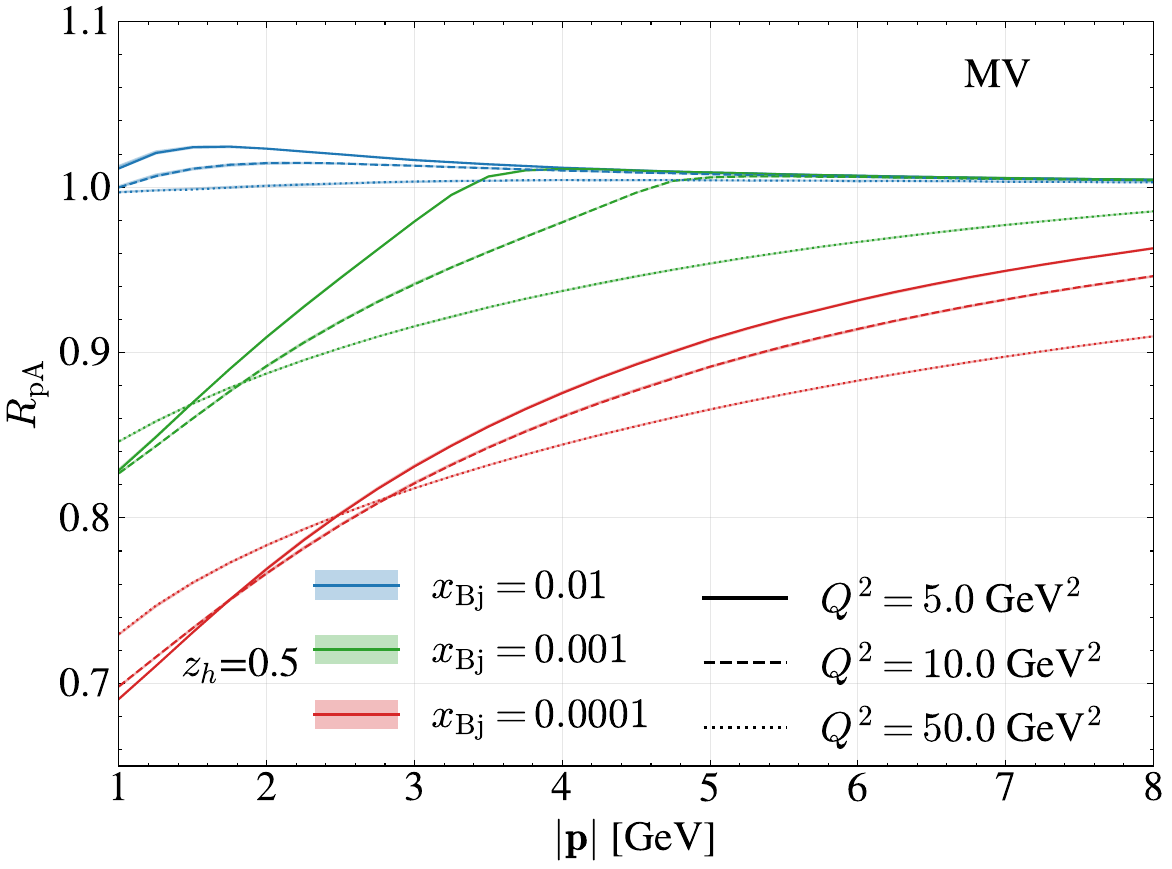}
\caption{MV}
\label{plot:Plot_SIDIS_nuclear_mod_EIC_z105_Q2_5_100_mb_MV}
\end{subfigure}
\hfill
\begin{subfigure}[t]{0.49\textwidth}
\centering
\includegraphics[width=\linewidth]{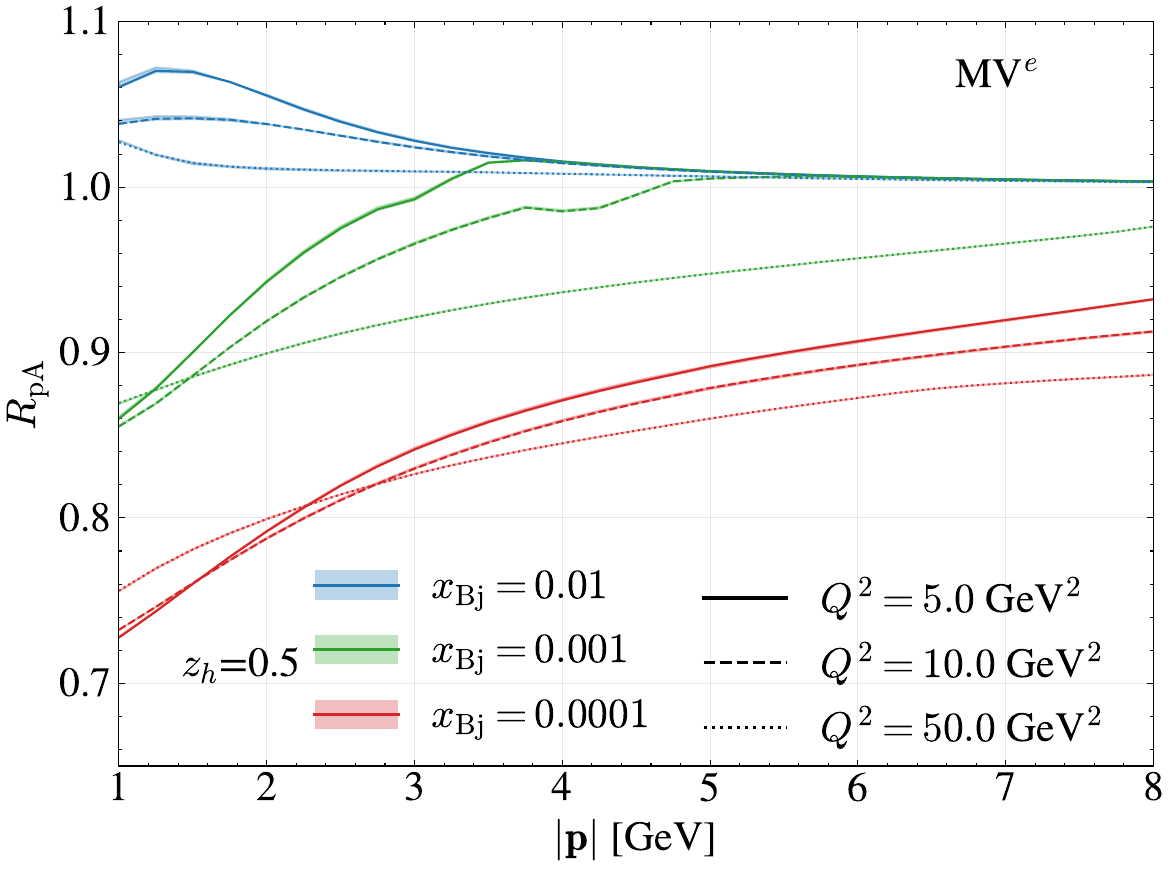}
\caption{MV$^e$}
\label{plot:Plot_SIDIS_nuclear_mod_EIC_z105_Q2_5_100_mb_MVe}
\end{subfigure}

\caption{Nuclear modification factor as a function of $|\pt|$ for selected values of $Q^2$ and $\xbj$ with fixed $z_h=0.5$. Results are shown for MV and MV$^e$ model fit initial conditions for dipole amplitudes.
}
\label{fig:Plot_SIDIS_nuclear_mod_EIC_z105_Q2_5_100_mb}
\end{figure*}
%

The larger saturation scale of the heavy nucleus typically suppresses the nuclear cross section compared to an incoherent superposition of $A$ nucleons. To quantify the nuclear suppression, which in our setup is due to gluon saturation, we compute the nuclear modification factor
\begin{equation}
    R_\mathrm{pA}^\mathrm{L,T} = \frac{{\dd \sigma^{\gamma^* + A \to h + X}_{\mathrm{L,T}}/(\dd[2]\pt \dd{z_h}) }}{A \times {\dd \sigma^{\gamma^* + p \to h + X}_{\mathrm{L,T}}/(\dd[2]\pt \dd{z_h})} }.
\end{equation}
In the dilute limit, and at the initial condition of the BK evolution, one obtains $R_\mathrm{pA}=1$, corresponding to no nuclear modification. Here $\mathrm{L}$ (longitudinal) and $\mathrm{T}$ (transverse) refer to the virtual photon polarization states, and by $R_\mathrm{pA}$ we denote the nuclear modification factor constructed from the sum of the transverse and longitudinal photon contributions to the cross sections. Nuclear modification factors presented in this Section also include the uncertainties originating from the fragmentation function scale choice, but it is
 negligible because this uncertainty almost completely cancels in the ratio. Uncertainties in the initial condition parametrization of dipole amplitude in the case of \mve model  fit are presented in  Appendix~\ref{append:A}. 

Nuclear modification factors in inclusive $\pi^\pm$ production as a function of pion transverse momenta $\pt$, computed using the MV and \mve dipole amplitude fits,
are shown in Fig.~\ref{fig:Plot_SIDIS_nuclear_mod_EIC_z105_Q2_5_100_mb}. Results are again calculated at fixed $z_h=0.5$, but the dependence on $z_h$ is found to be weak.
A small Cronin enhancement is visible at low-$\pt$ especially when the MV$^e$ fit is used,  and by construction $R_\mathrm{pA}$ approaches unity at high $|\pt|$.
Significant nuclear suppression is predicted for $x\lesssim 10^{-3}$, and the suppression becomes stronger with decreasing $x$ as the nuclear saturation scale $Q_s$ increases.
The dependence on the photon virtuality $Q^2$ is relatively weak, which is also expected in the TMD limit $|\pt|^2\ll Q^2$~\cite{Marquet:2009ca}.
The $\pt$ dependence of the nuclear modification factor becomes  milder towards higher $Q^2$. Because of this, and the fact that $x_g$ decreases with increasing $Q^2$‚  we find somewhat stronger suppression  at intermediate $|\pt|$ for larger values of $Q^2$.
The predicted dependence on $\pt$ and $\xbj$, as well as the overall magnitude of the nuclear suppression, is qualitatively consistent with the results of Ref.~\cite{Lappi:2013zma,Albacete:2003iq}, where an equivalent CGC setup has been applied to inclusive hadron production in proton-nucleus collisions. 

The $R_\mathrm{pA}$ predictions obtained using the MV and \mve dipole amplitude fits shows noticeable differences. The \mve fit predicts slightly  stronger suppression (smaller $R_\mathrm{pA}$) at large $\pt$ than the MV model, and exhibits a slightly weaker dependence on $\pt$. Furthermore,  the \mve parametrization leads to a  more pronounced Cronin peak ($R_\mathrm{pA}>1$ at small $|\pt|$) at $\xbj=0.01$.
Given that both dipole model fits provide a good description of the inclusive structure function data, observed differences suggests that future $R_{pA}$ measurements at the EIC, which are sensitive to the momentum space dipole, could provide some complementary constraints on the initial condition of the BK equation.

Next we study the nuclear modification factor separately for transverse and longitudinal photons. Although such measurements are experimentally significantly more challenging than the total SIDIS cross section considered above, it is still instructive to determine if even stronger sensitivity to the BK equation initial condition could be achieved with such data.

\begin{figure*}[t]
\centering

\begin{subfigure}[t]{0.49\textwidth}
\centering
\includegraphics[width=\linewidth]{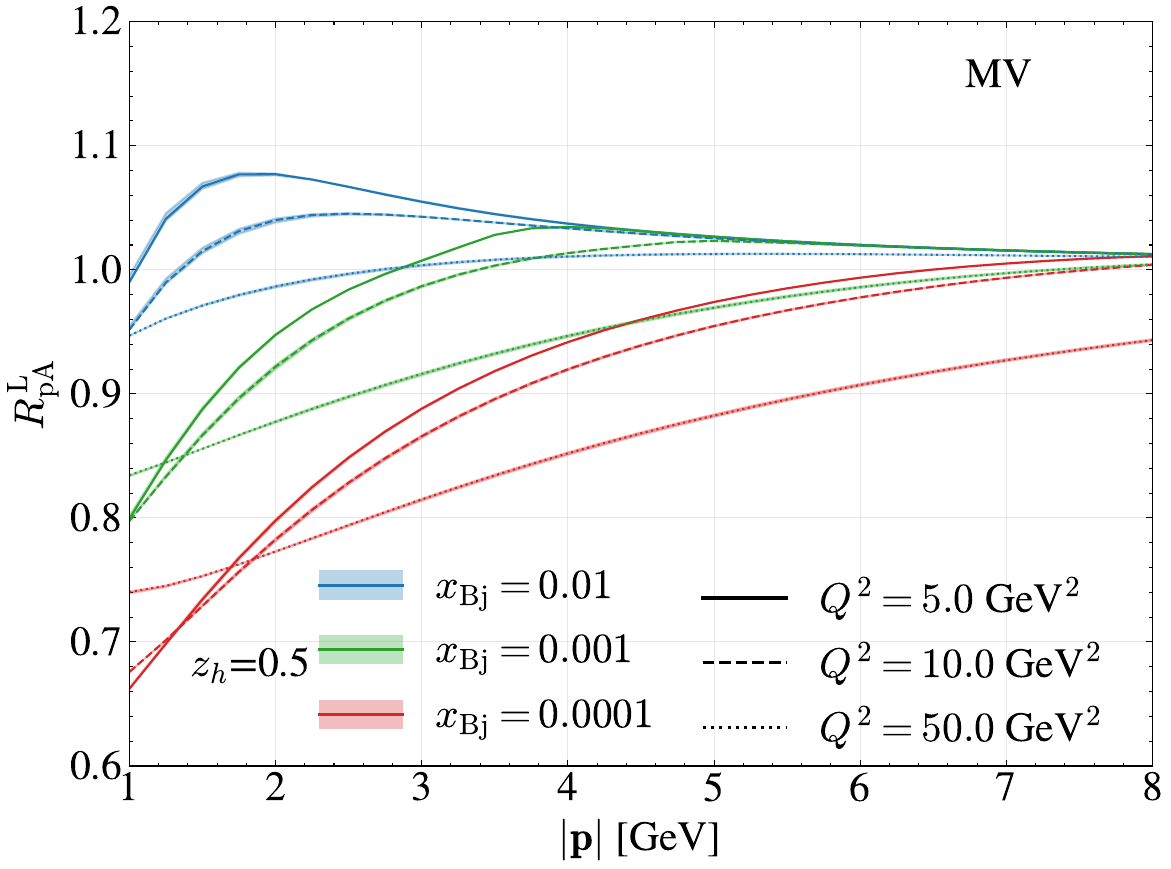}
\caption{MV}
\label{plot:Plot_SIDIS_nuclear_mod_L_EIC_z105_Q2_5_100_mb_MV}
\end{subfigure}
\hfill
\begin{subfigure}[t]{0.49\textwidth}
\centering
\includegraphics[width=\linewidth]{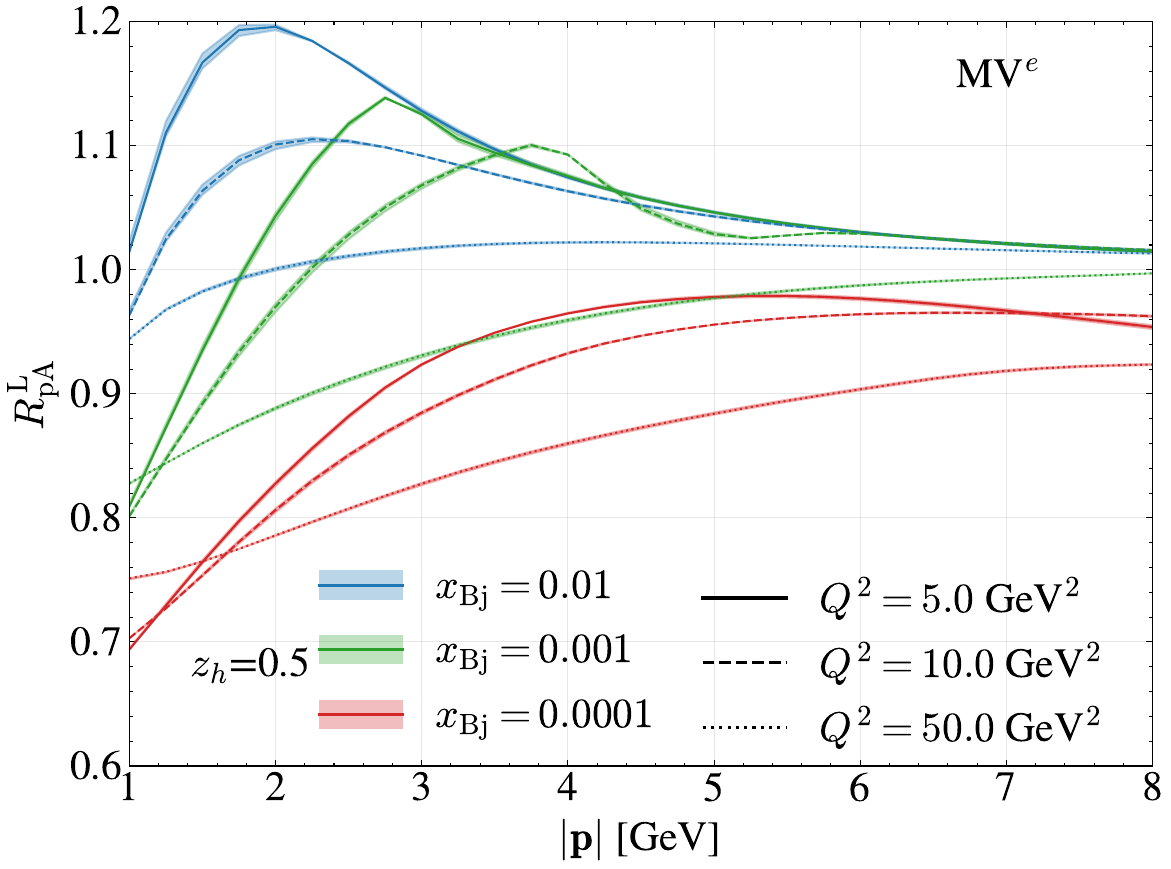}
\caption{MV$^e$}
\label{plot:Plot_SIDIS_nuclear_mod_L_EIC_z105_Q2_5_100_mb_\mve}
\end{subfigure}

\caption{Nuclear modification factor as a function of pion transverse momenta $|\pt|$, for longitudinally polarized photon at selected values of  $Q^2$ and  $\xbj$ with fixed $z_h=0.5$. Results are shown for MV and MV$^e$ model fit initial conditions for the dipole amplitudes. }
\label{fig:Plot_SIDIS_nuclear_mod_L_EIC_z105_Q2_5_100_mb}
\end{figure*}

The nuclear modification factor $R_\mathrm{pA}^\mathrm{L}$  for longitudinally polarized photons as a function of hadron $\pt$, at fixed $z_h = 0.5$, is shown in Fig.~\ref{fig:Plot_SIDIS_nuclear_mod_L_EIC_z105_Q2_5_100_mb}. 
The $R_\mathrm{pA}^\mathrm{L}$ depends more strongly on the pion transverse momentum $\pt$ compared to $R_\mathrm{pA}$, and differences between the MV and \mve parametrizations are substantially enhanced. 
This can be understood from the fact that a longitudinal photon typically splits into dipoles of size $\sim 1/Q$, whereas the corresponding dipole size distribution with transverse photons (which dominate the total cross section) is considerably broader. A much stronger Cronin peak is also predicted when the \mve dipole is used. 
The BK evolution is known to wash out the Cronin enhancement towards small-$\xbj$~\cite{Albacete:2003iq}, but a significantly longer evolution is required in the \mve case in order to remove this enhancement compared to the calculation with the MV dipole.

\begin{figure*}[t]
\centering

\begin{subfigure}[t]{0.49\textwidth}
\centering
\includegraphics[width=\linewidth]{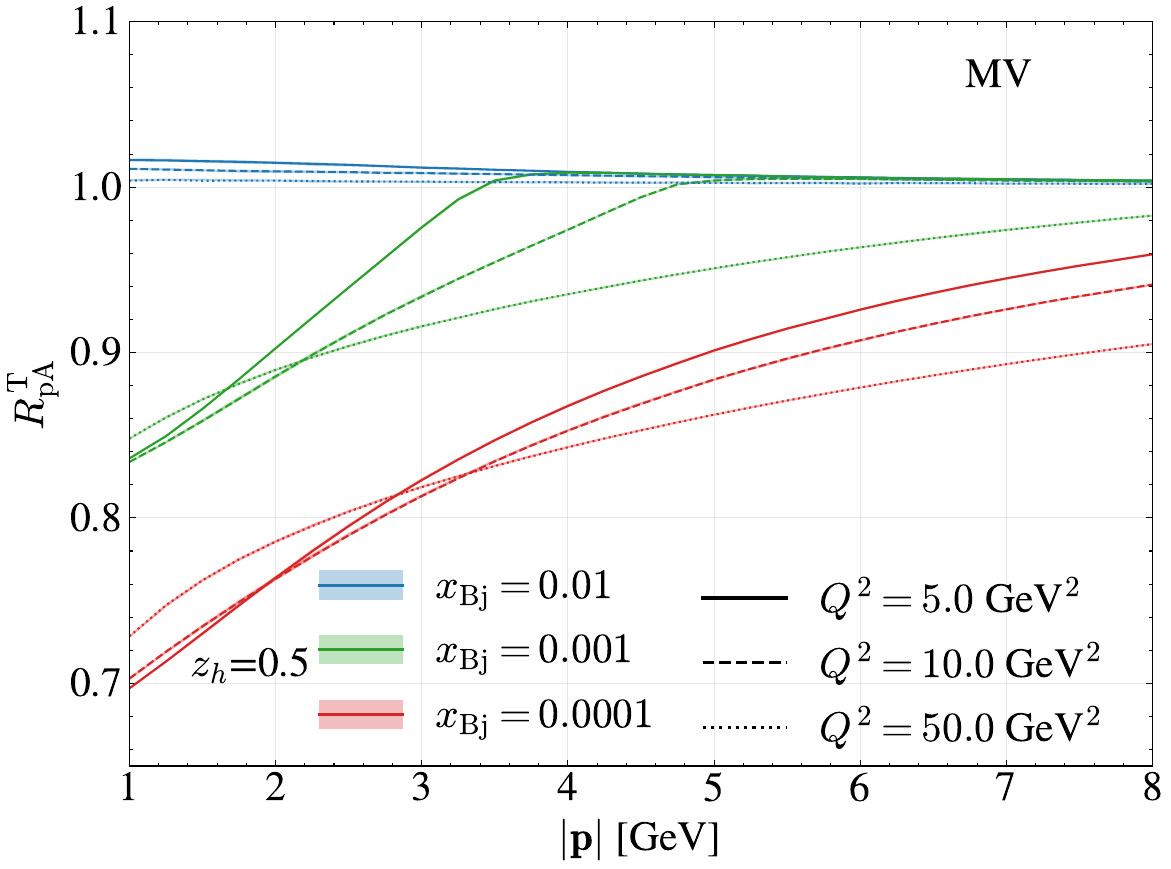}
\caption{MV}
\label{plot:Plot_SIDIS_nuclear_mod_T_EIC_z105_Q2_5_100_mb_MV}
\end{subfigure}
\hfill
\begin{subfigure}[t]{0.49\textwidth}
\centering
\includegraphics[width=\linewidth]{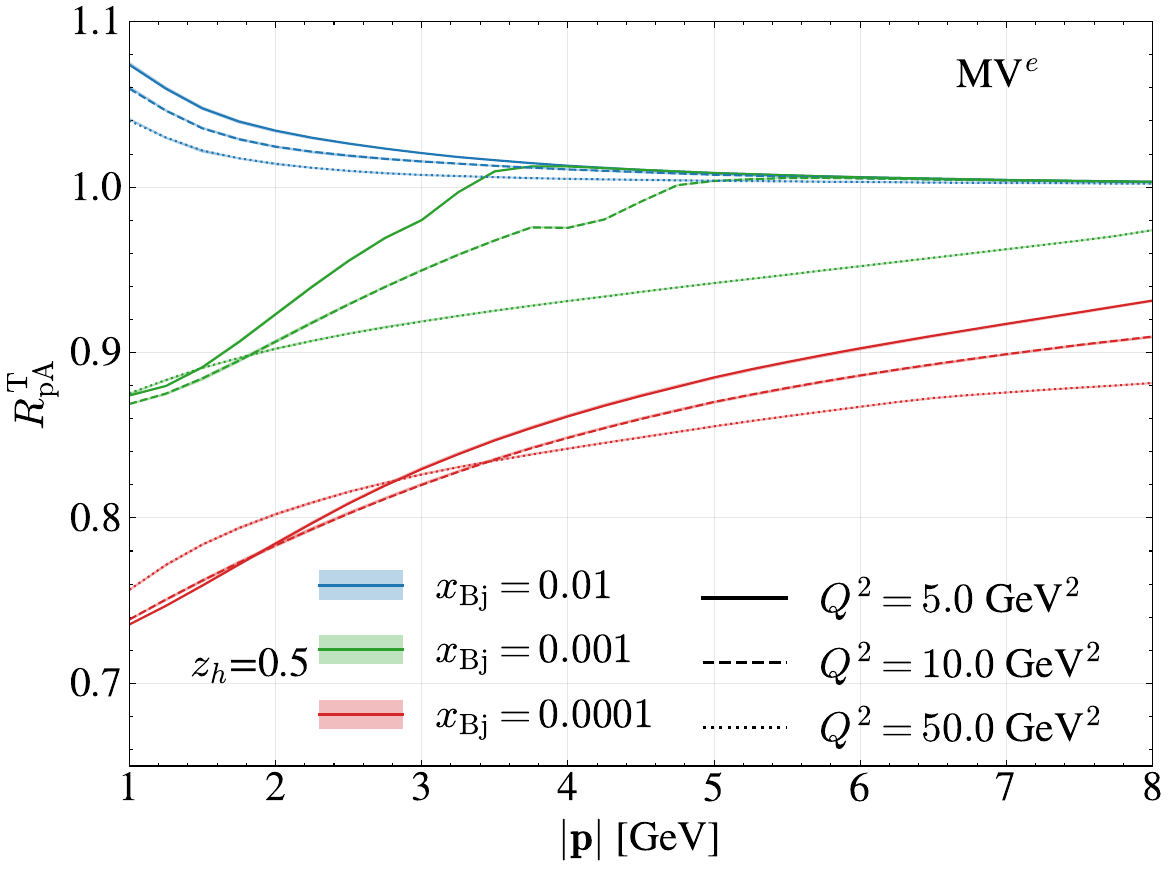}
\caption{MV$^e$}
\label{plot:Plot_SIDISplots/plot_expanded/Plot_SIDIS_nuclear_mod_T_EIC_z105_Q2_5_100_mb_MV_T_expanded.pdf_nuclear_mod_T_EIC_z105_Q2_5_100_mb_\mve}
\end{subfigure}

\caption{Nuclear modification factor as a function of pion transverse momenta $|\pt|$, for transverse polarization of photon at selected values of  $Q^2$ and  $\xbj$ with fixed $z_h=0.5$. Results are shown for MV and MV$^e$ model fit initial conditions for the dipole amplitudes.}
\label{fig:Plot_SIDIS_nuclear_mod_T_EIC_z105_Q2_5_100_mb}
\end{figure*}

Fig.~\ref{fig:Plot_SIDIS_nuclear_mod_T_EIC_z105_Q2_5_100_mb}  shows the nuclear modification factor for transversely polarized photons as a function of pion $\pt$. 
Since the transverse contribution dominates the cross section in this kinematic regime, $R_\mathrm{pA}^\mathrm{T}$ exhibits behavior similar to that of the total nuclear modification factor $R_\mathrm{pA}$ shown in Fig.~\ref{fig:Plot_SIDIS_nuclear_mod_EIC_z105_Q2_5_100_mb}. The main difference is that the Cronin peaks at $\xbj=0.01$ obtained with the MV$^e$ model are shifted to smaller $|\pt|$.

A Cronin enhancement in DIS that is washed out by BK evolution, together with nuclear suppression at small $x$, was previously reported in Ref.~\cite{Iancu:2020jch} for quark (jet) production at very forward rapidities, corresponding to $z_q \to 1$. In this regime, $z_q(1-z_q)Q^2 \ll Q_s^2$, and the photon predominantly splits into large dipoles that interact strongly in the black disk regime.
Although we consider a different process, namely inclusive hadron production in realistic HERA/EIC kinematics, and employ a different construction of the dipole-nucleus scattering amplitude~\eqref{eq:dipole_nucleus} than Ref.~\cite{Iancu:2020jch} where a simple $A^{1/3}$ scaling of the saturation scale was assumed, our results exhibit the same qualitative behavior: a Cronin enhancement at moderately large $x$ and substantial nuclear suppression at small $x$. 

\begin{figure*}[t]
\centering

\begin{subfigure}[t]{0.49\textwidth}
\centering
\includegraphics[width=\linewidth]{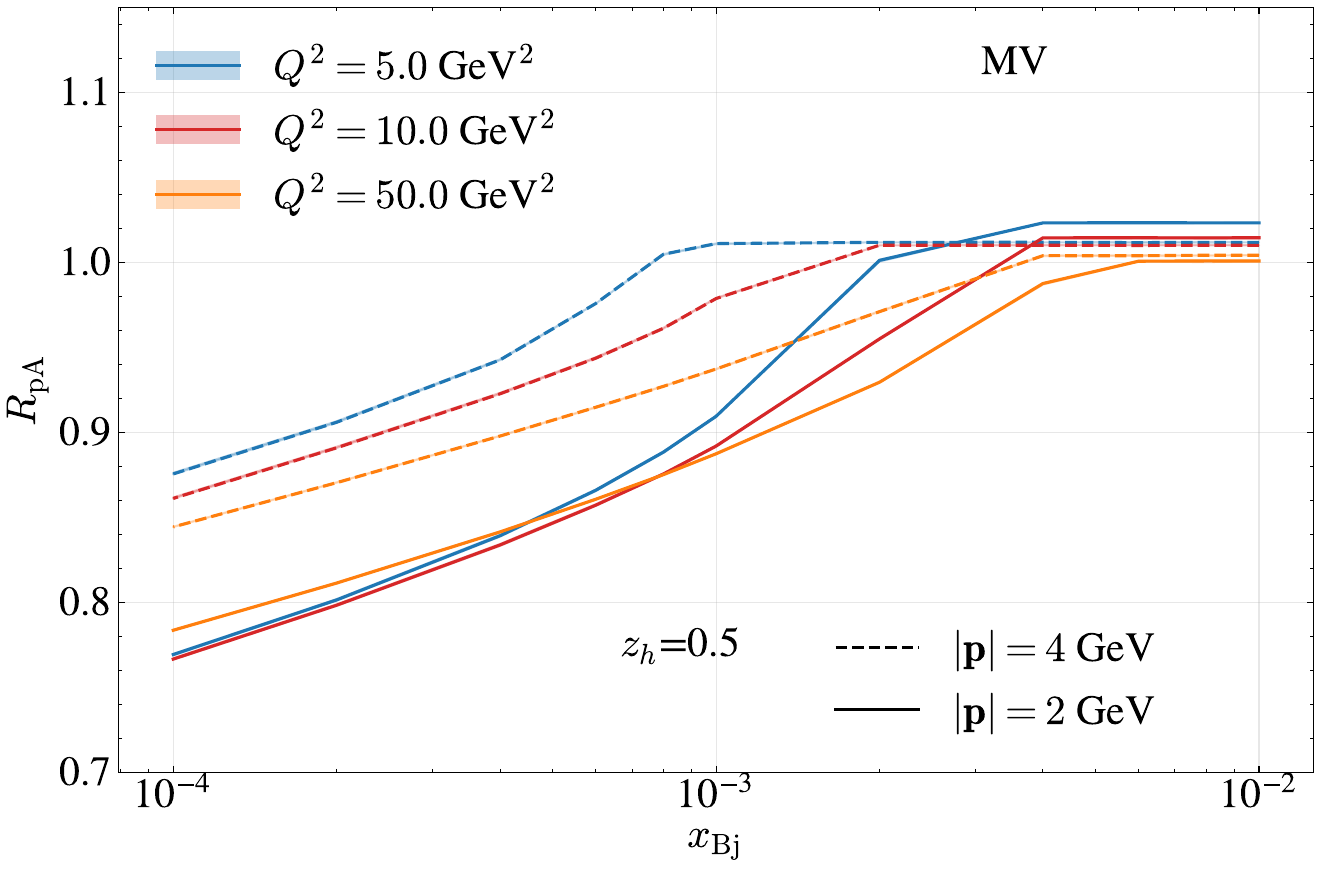}
\caption{MV}
\label{plot:Plot_SIDIS_nuclear_mod_EIC_R_vs_xbj_z105_Pper_2_8_MV}
\end{subfigure}
\hfill
\begin{subfigure}[t]{0.49\textwidth}
\centering
\includegraphics[width=\linewidth]{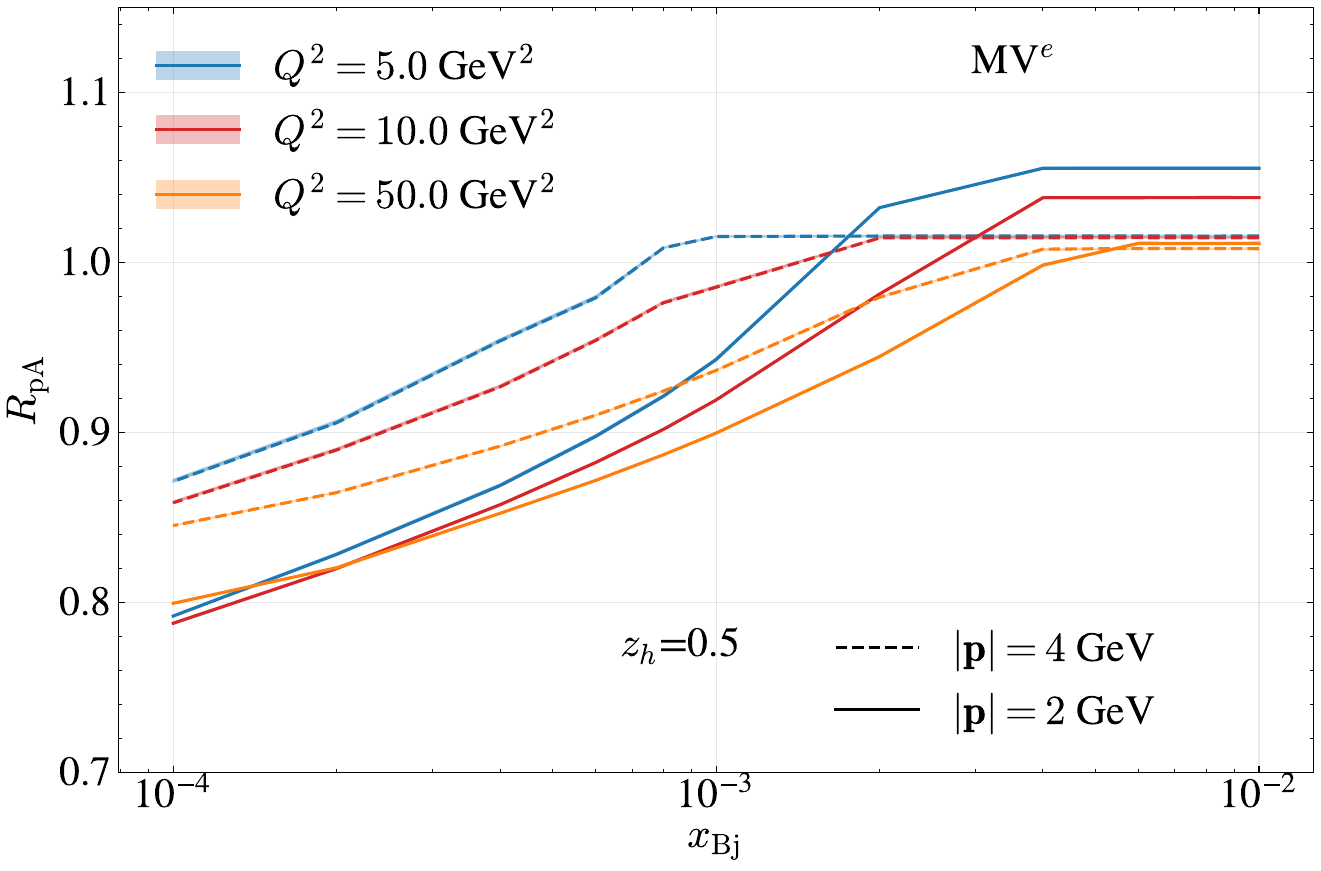}
\caption{MV$^e$}
\label{plot:/Plot_SIDIS_nuclear_mod_T_EIC_z105_Q2_5_100_mb_\mve}
\end{subfigure}

\caption{Nuclear modification factor as a function of $\xbj$ for selected values of photon virtuality $Q^2$ and transverse momenta of produced pion $|\pt|$ with fixed $z_h = 0.5$. The results are obtained using MV and MV$^e$ model initial conditions for dipole amplitudes.
In the regime where $R_\mathrm{pA}$ becomes independent of $\xbj$ our results are sensitive to the extrapolation of the framework beyond its range of validity $x_g<0.01$.
}
\label{fig:Plot_SIDIS_nuclear_mod_EIC_R_vs_xbj_z105_Pper_2_8}
\end{figure*}

Finally we study the nuclear modification factor as a function of $\xbj$ at fixed $z_h = 0.5$ and $|\pt|=2\,\mathrm{GeV}$ and $|\pt|=4\,\mathrm{GeV}$. The results are shown in Fig.~\ref{fig:Plot_SIDIS_nuclear_mod_EIC_R_vs_xbj_z105_Pper_2_8}.
We again find that the nuclear suppression typically becomes stronger towards smaller $\xbj$ (corresponding to a larger saturation scale $Q_s^2$) and lower $|\pt|$. The notable exception is the \mve parametrization at low $Q^2$ and $|\pt|$, where the Cronin enhancement ($R_\mathrm{pA}>1$) is visible. 
Since the dipole amplitude is frozen at the initial condition at $x_g>x_0=0.01$, the $R_\mathrm{pA}$ becomes $\xbj$-independent for $\xbj \gtrsim 0.003$. As noted above, this behavior is an artifact of our numerical implementation and indicates that in this regime our predictions become sensitive to the extrapolation of our setup outside its range of validity $x_g<0.01$.

%
%

\section{Conclusions}
\label{sec:conclusions}


We have calculated single inclusive hadron production in Deep Inelastic Scattering within the Color Glass Condensate framework. The leading order setup constrained by HERA structure function data is validated against HERA SIDIS measurements, and a reasonably good agreement is found especially at low hadron transverse momentum.
The setup is then applied to make predictions for electron-nucleus scattering. We predict the nuclear modification factor due to gluon saturation effects that can be measured at the EIC. Compared to DIS structure functions, more differential measurements as a function of hadron momentum are possible, which can allow for more detailed studies of gluon saturation.

We predict a significant nuclear suppression in inclusive pion production in EIC kinematics, with the suppression becoming stronger towards smaller $\xbj$, where saturation effects are more pronounced, and lower hadron transverse momentum. At moderate $\xbj \sim 10^{-2}$, a Cronin peak is predicted to be visible, and it is significantly more pronounced in the case of longitudinally polarized photons. This Cronin enhancement disappears as a consequence of the BK evolution towards smaller-$\xbj$. These features are similar to what has been predicted for single inclusive hadron production in proton-nucleus collisions using a similar CGC framework~\cite{Albacete:2003iq,Lappi:2013zma}.

We present results obtained using two different initial conditions for the BK evolution, both fitted to HERA structure function data. 
Despite providing a comparable description of the structure function measurements, the two parametrizations lead to numerically 
  significant differences in the predicted nuclear suppression factors, affecting both the overall level  of the nuclear suppression and  the magnitude of the Cronin enhancement. This suggests that future nuclear-DIS  measurements can provide  additional constraints on the initial condition BK equation when incorporated into global analyses.

Because the phase space of the unmeasured quark is integrated over analytically, the longitudinal momentum fraction of the target nucleon transferred to the produced system has to be estimated. When the estimated invariant mass of the final state $|q\bar q\rangle$ system is large, one is sensitive to the dipole-target scattering amplitude at $x_g>0.01$, i.e. at higher-$x_g$ than the one where the initial condition of the BK evolution is parametrized. Consequently, the results obtained in this work become sensitive to the chosen extrapolation scheme at $\xbj \gtrsim 0.003$. If DIS structure function fits applying a similar choice for the evolution rapidity and a realistic extrapolation to the larger-$x_g$ regime become available in the future, based e.g. on Ref.~\cite{Bertilsson:2026vtu}, the dependence on this scheme choice will likely be reduced.  

We also present a numerically stable method to evaluate the dipole amplitude in momentum space, which allows us to compute  cross section for large final state hadron transverse momentum $\pt$.

We quantify two types of uncertainties in our studies.
Fragmentation function uncertainty is estimated by varying the scale choice, which results in moderate uncertainty at the cross section level, but this uncertainty almost completely cancels in cross section ratios like the nuclear suppression factor. On the other hand, uncertainties in the BK initial condition (quantified in Appendix~\ref{append:A}), are still visible in the nuclear suppression factor. This also indicates the potential of future EIC data on constraining the non-perturbative input for the small-$x$ BK evolution. 


The accuracy of our results can be systematically improved. The SIDIS cross section at NLO accuracy has been recently derived~\cite{Bergabo:2024ivx,Bergabo:2022zhe, Altinoluk:2025dwd,Caucal:2024cdq}, and when coupled with NLO BK evolution~\cite{Balitsky:2007feb} and NLO dipole amplitude~\cite{Casuga:2026xxt,Hanninen:2022gje}, could be applied to EIC phenomenology.
Similarly, one can go beyond the high-energy limit and systematically include subeikonal corrections, which have been derived for SIDIS in Ref.~\cite{Altinoluk:2025ang}.
Sudakov effects can also be numerically important and could be included~\cite{Altinoluk:2024vgg}.


\subsection*{Data Availability}
The results shown in this work are obtained using the publicly available codes~\cite{sidis_lo_code}.

\begin{acknowledgements}
We thank L. Huhta, F. Salazar, S. Tiwari and A. Vilenius for discussions. This work was supported by the Research Council of Finland, the Centre of Excellence in Quark Matter, and projects 338263 and 359902, and by the European Research Council (ERC, grant agreements  No. ERC-2023-101123801 GlueSatLight and ERC-2018-ADG-835105 YoctoLHC).
C.C acknowledges the support of the Vilho, Yrjö and Kalle Väisälä Foundation.
Computing resources from CSC – IT Center for Science in Finland and the Finnish Grid and Cloud Infrastructure (persistent identifier \texttt{urn:nbn:fi:research-infras-2016072533}) were used in this work.
The content of this article does not reflect the official opinion of the European Union and responsibility for the information and views expressed therein lies entirely with the authors.
\end{acknowledgements}

\appendix

\section{Uncertainty estimate from the BK initial condition}
\label{append:A}
The uncertainty band shown in Figs.~\ref{plot:Plot_FF_Scale_Uncertainty_HERA_3Tables_gP_HERA_VS_MVe_VS_MV} to~\ref{fig:Plot_SIDIS_nuclear_mod_EIC_R_vs_xbj_z105_Pper_2_8} 
corresponds to the  fragmentation function scale uncertainty, estimated by varying the scale by a factor $2$. In this Appendix, on the other hand, we propagate uncertainties in the dipole-target scattering amplitude to the SIDIS cross section.

The \mve dipole amplitude fit from Ref.~\cite{Casuga:2023dcf} provides a set of BK initial conditions sampled from the posterior distribution obtained in the fit to DIS structure function data. The uncertainty of a predicted observable is estimated by evaluating it for each posterior sample and computing the variance of the resulting predictions. No such uncertainty estimates are available for the MV dipole fit~\cite{Lappi:2013zma}.
 \begin{figure*}[t]
 \centering

 \begin{subfigure}{0.49\textwidth}
 \centering
 \includegraphics[width=\linewidth]{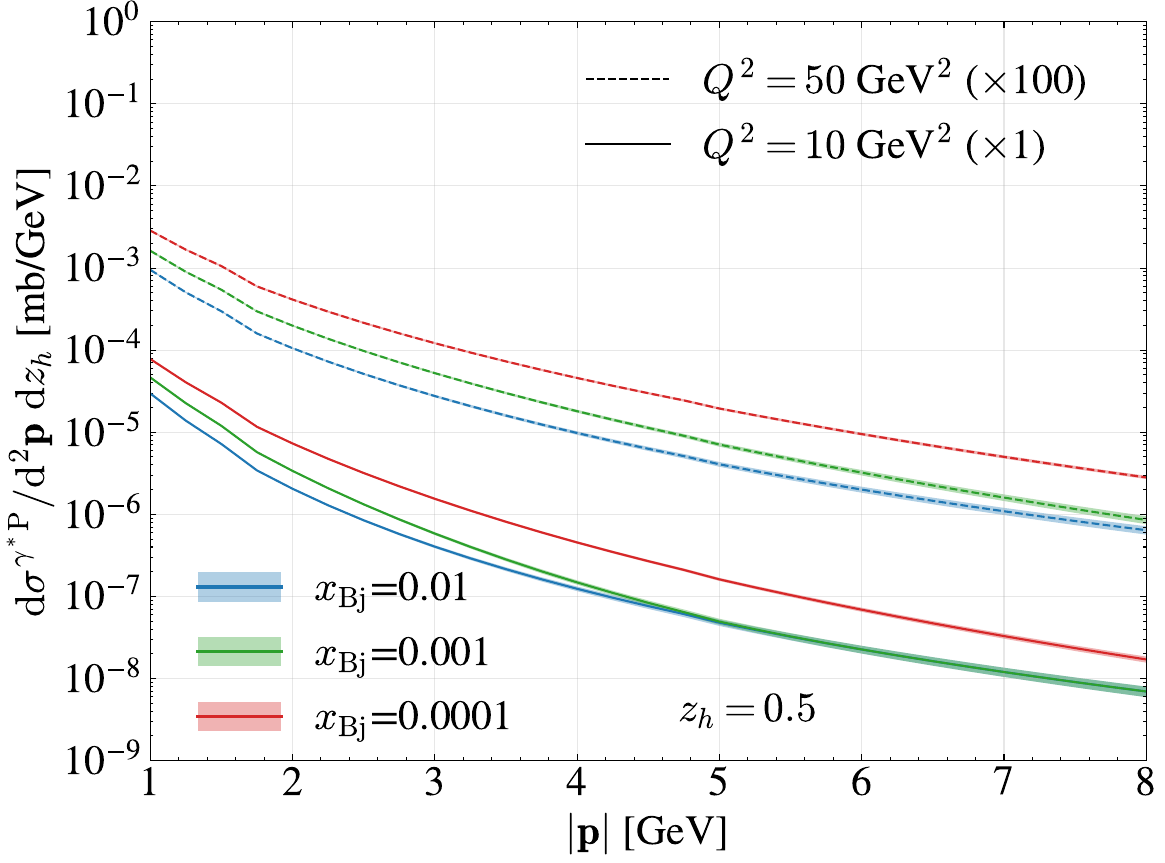}
 \caption{$\gamma^*$-P}
 \label{plot:Plot_posterior_sample_uncertainty_sigma_Vs_Pper_EIC_gP_z10.5}
 \end{subfigure}
 \hfill
 \begin{subfigure}{0.49\textwidth}
 \centering
 \includegraphics[width=\linewidth]{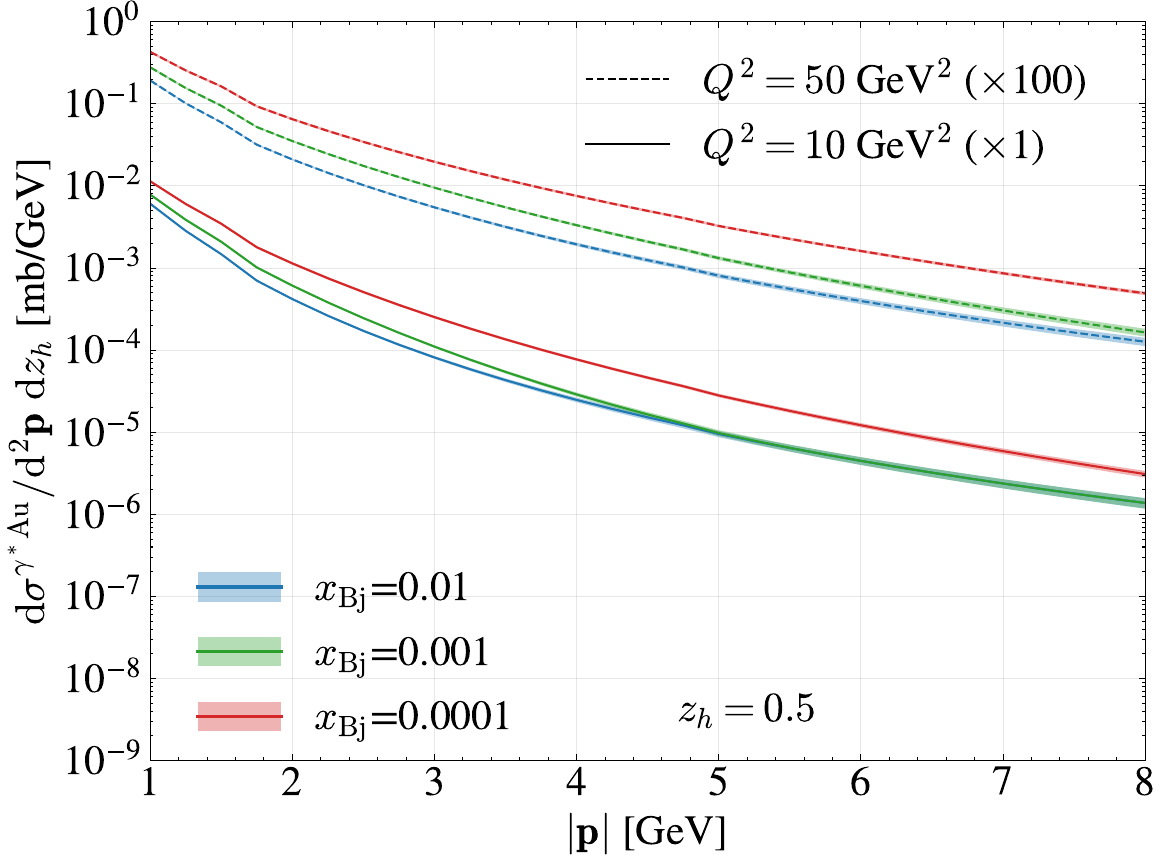}
 \caption{$\gamma^*$-Au}
 \label{plot:Plot_posterior_sample_uncertainty_sigma_Vs_Pper_EIC_gA_z10.5}
 \end{subfigure}

 \caption{Differential cross section for $\gamma^*$-proton and $\gamma^*$-Au scattering as a function of $|\pt|$, for selected values of $Q^2$ and $\xbj$ with fixed $z_h = 0.5$. The results are obtained by averaging over posterior samples from the dipole amplitudes fitted using MV$^e$ model initial conditions. For clarity, the results at $Q^2$ = 50 GeV$^2$ are multiplied by 100.
 }
 \label{fig:Plot_posterior_sample_uncertainty_sigma_Vs_Pper}

 \end{figure*}
In Fig.~\ref{fig:Plot_posterior_sample_uncertainty_sigma_Vs_Pper} we show the  differential SIDIS cross section for $\gamma^*$-proton and $\gamma^*$-Au scattering as a function of final state charged pion transverse momentum $\pt$. The central line represents the mean cross section obtained by calculating the cross section using different posterior samples for the dipole amplitude, while the uncertainty band corresponds to two standard deviations. This uncertainty is in practice small, at most $\sim 15\%$
at high $|\pt|$. This relative uncertainty is similar for both proton and nuclear targets.

\begin{figure}[t]
     \centering
     \includegraphics[width=.5\textwidth]{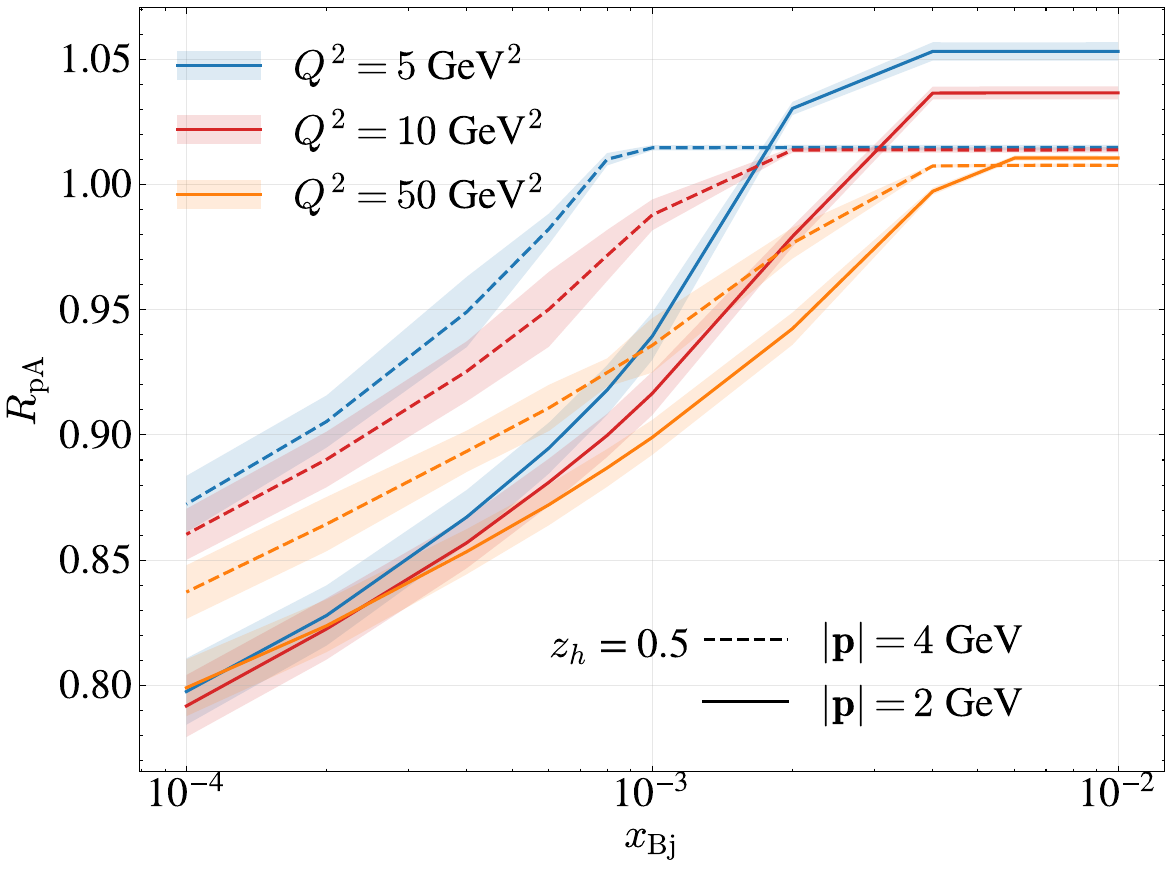}
     \caption{Nuclear modification factor as a function of  $\xbj$ for selected values of $Q^2$ and $|\pt|$ at fixed $z_h=0.5$; the central lines represent mean over posterior distributions of dipole amplitudes, while the bands represent the corresponding $\pm2\sigma$ uncertainty. In the regime where $R_\mathrm{pA}$ becomes independent of $\xbj$ our results are sensitive to the extrapolation of the framework beyond its range of validity $x_g<0.01$.}
     \label{fig:Plot_Posterior_sample_mean_SIDIS_nuclear_mod_EIC_R_vs_xbj_z105_Pper_2_8_MVe}
 \end{figure}

\begin{figure}[t]
    \centering
    \includegraphics[width=.5\textwidth]{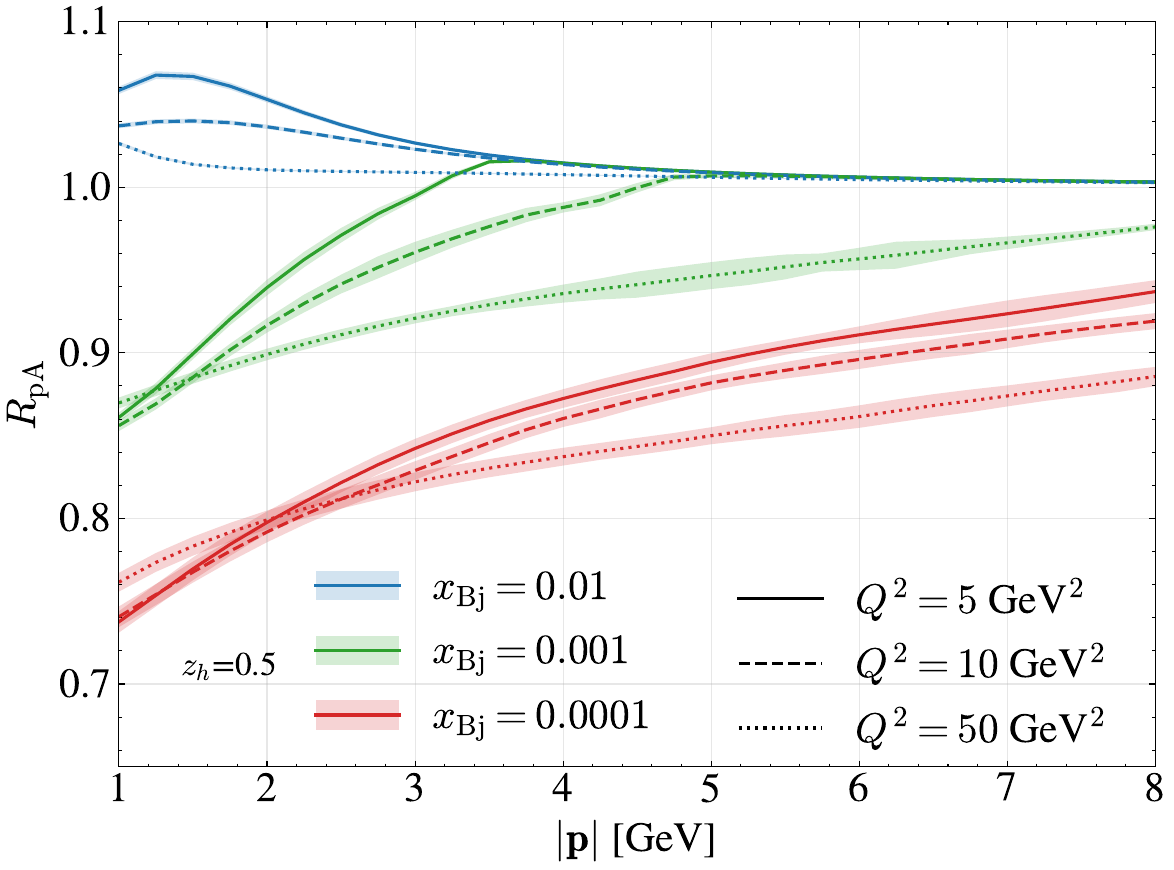}
     \caption{Nuclear modification factor as a function of $|\pt|$ for selected values of $Q^2$ and  $\xbj$ at fixed $z_h=0.5$. The central lines represent the posterior mean of nuclear modification factors obtained using the MV$^e$ model initial conditions for the dipole amplitudes, while bands represent corresponding $\pm2\sigma$ uncertainty intervals.}
     \label{plot:Plot_Posterior sample_mean_SIDIS_nuclear_mod_EIC_z105_Q2_5_100_mb_MVe}
 \end{figure}

\begin{figure*}[!htbp]
 \centering

 \begin{subfigure}{0.49\textwidth}
 \centering
 \includegraphics[width=\linewidth]{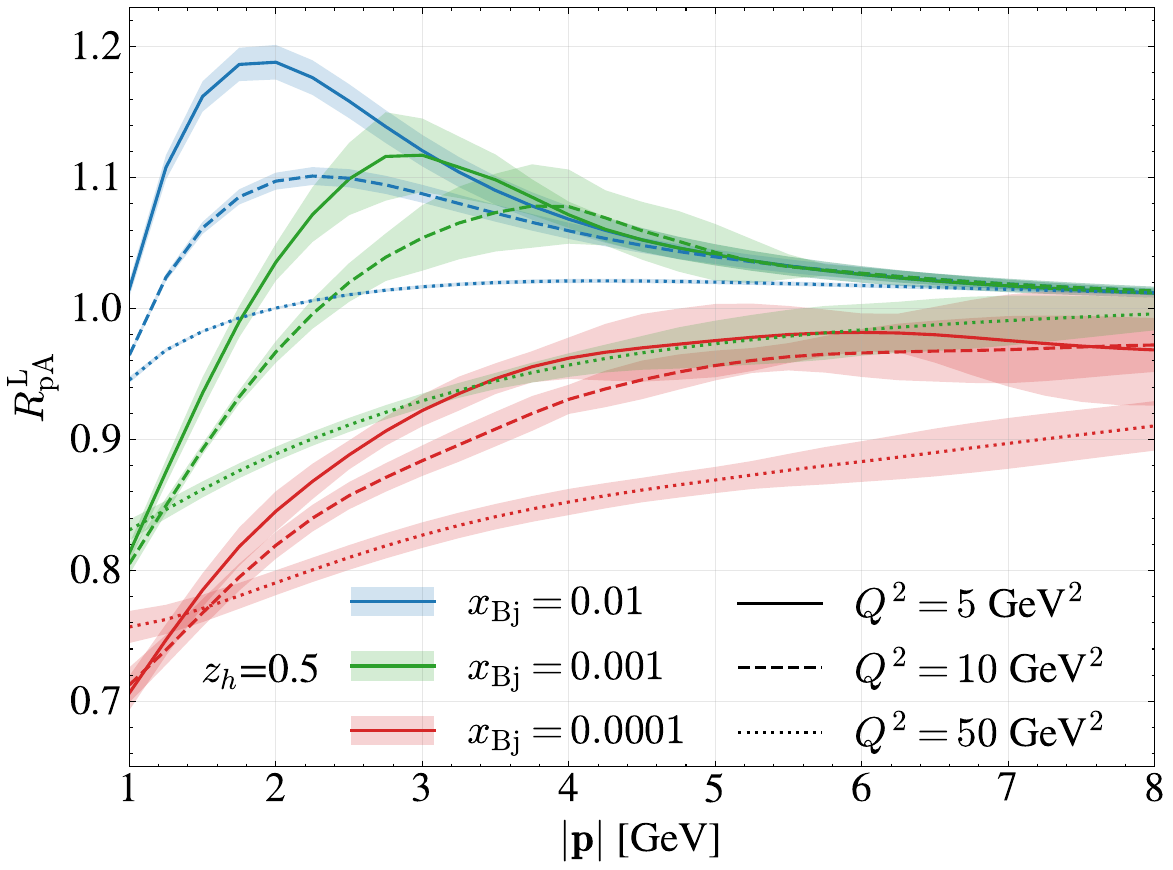}
 \caption{Longitudinal}
 \label{plot:SIDIS_L_nuclear_mod_posterior_sample_Vs_pper}
 \end{subfigure}
 \hfill
 \begin{subfigure}{0.49\textwidth}
 \centering
 \includegraphics[width=\linewidth]{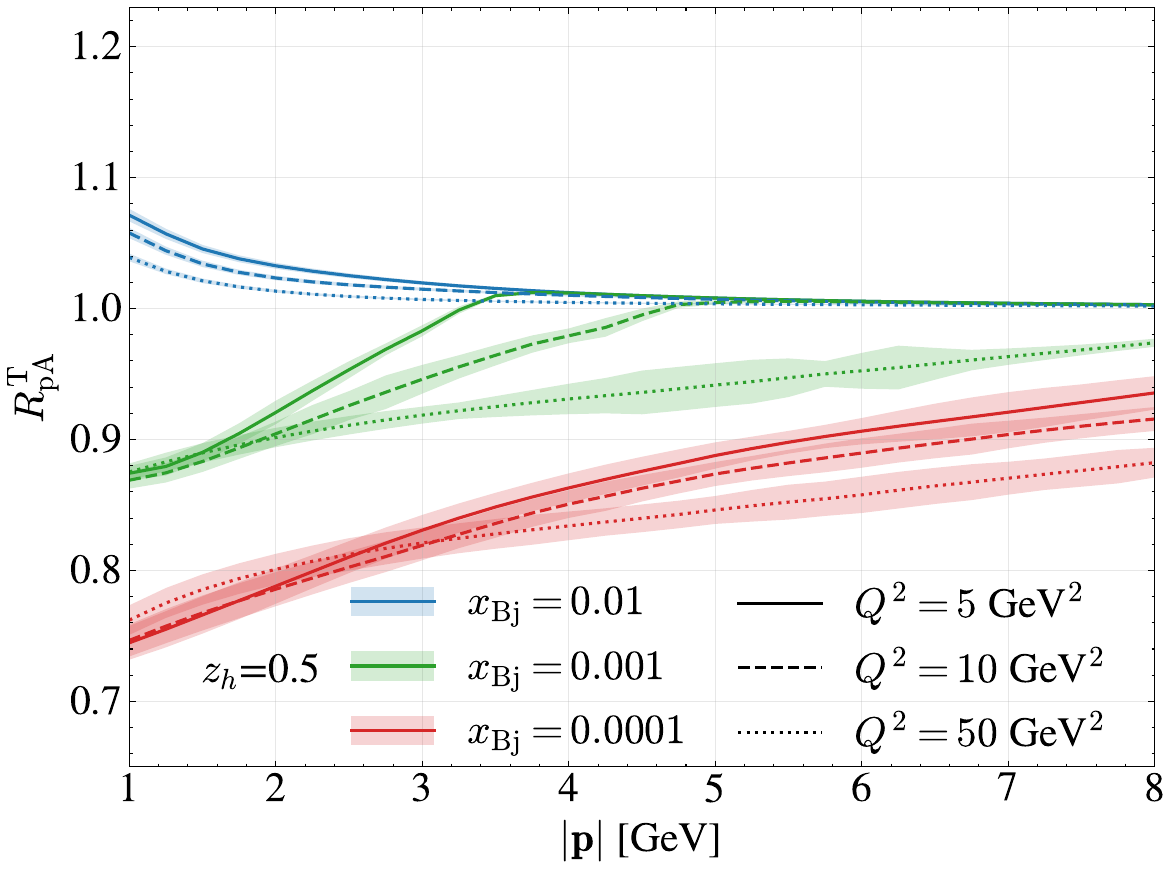}
 \caption{Transverse}
 \label{plot:SIDIS_T_nuclear_mod_posterior_sample_Vs_per}
 \end{subfigure}

 \caption{Nuclear modification factor as a function of $|\pt|$ for selected values of $Q^2$ and  $\xbj$ at fixed $z_h=0.5$ separately for longitudinal and transverse polarizations of incoming photon. The central lines represent the posterior mean of nuclear modification factors obtained using the MV$^e$ model initial conditions for the dipole amplitudes, while bands represent corresponding $\pm2\sigma$ uncertainty intervals.}
\label{fig:SIDIS_nuclear_mod_EIC_posterior_sample_Vs_Pper}
 \end{figure*}

Next we propagate the BK initial condition uncertainty to the nuclear suppression factor. First, in Fig.~\ref{fig:Plot_Posterior_sample_mean_SIDIS_nuclear_mod_EIC_R_vs_xbj_z105_Pper_2_8_MVe} we show $R_\mathrm{pA}$ as a function of $\xbj$ for fixed values of $\pt$. 
The obtained uncertainty (again corresponds to two standard deviations) grows slightly towards smaller $x$, and is  systematically larger than the fragmentation function uncertainty shown in Fig.~\ref{fig:Plot_SIDIS_nuclear_mod_EIC_R_vs_xbj_z105_Pper_2_8}. 
The nuclear modification factor as a function of the hadron transverse momentum $\pt$ is shown in  Fig.~\ref{plot:Plot_Posterior sample_mean_SIDIS_nuclear_mod_EIC_z105_Q2_5_100_mb_MVe}. The uncertainty is moderate and almost independent of $\pt$. We also note that the central values of $R_\mathrm{pA}$ obtained by averaging over posterior samples are, in practice, identical to those obtained  using the parametrization
in Sec.~\ref{sec:results} shown in Table.~\ref{tab:ic_parameters}.
Finally, Fig.~\ref{fig:SIDIS_nuclear_mod_EIC_posterior_sample_Vs_Pper} shows the uncertainty in the nuclear modification factor separately for longitudinally and transversely polarized virtual photons, with the uncertainty being slightly larger for the longitudinal contribution.

We conclude that the uncertainties associated with the non-perturbative BK evolution initial condition are at most moderate. In contrast, the systematic model uncertainty, estimated here from the difference between the \mve and MV initial condition parametrizations, is larger. This indicates that the choice of parametrization in the BK fits is likely the dominant contributor to the overall model uncertainty.
\section{Choice of scale for BK evolution of dipole amplitude}
\label{append:evolutionscale}

In the applied fits, the initial condition for the BK evolution is parametrized at the evolution rapidity $Y_0=\ln 1/x_0$, with $x_0=0.01$. When calculating the SIDIS cross section, the dipole amplitude is obtained by solving the BK equation up to evolution rapidity $Y=\ln 1/x_g$. As discussed in Sec.~\ref{sec:setup}, the choice~\eqref{eq:xg} for $x_g$ is not unique. Although this choice formally corresponds to a higher-order effect, it can  be numerically significant in practice.

The main results of this work use the $x_g$ estimate suggested in Ref.~\cite{Iancu:2020jch}, given in Eq.~\eqref{eq:xg}, with the dipole amplitude frozen to its initial condition for $x_g>x_0$. In this Appendix, we quantify the sensitivity of our main results, namely the nuclear modification factors predicted for the EIC,  to this choice of the BK evolution range. 

\begin{figure*}[!htbp]
\centering

\begin{subfigure}[t]{0.49\textwidth}
\centering
\includegraphics[width=\linewidth]{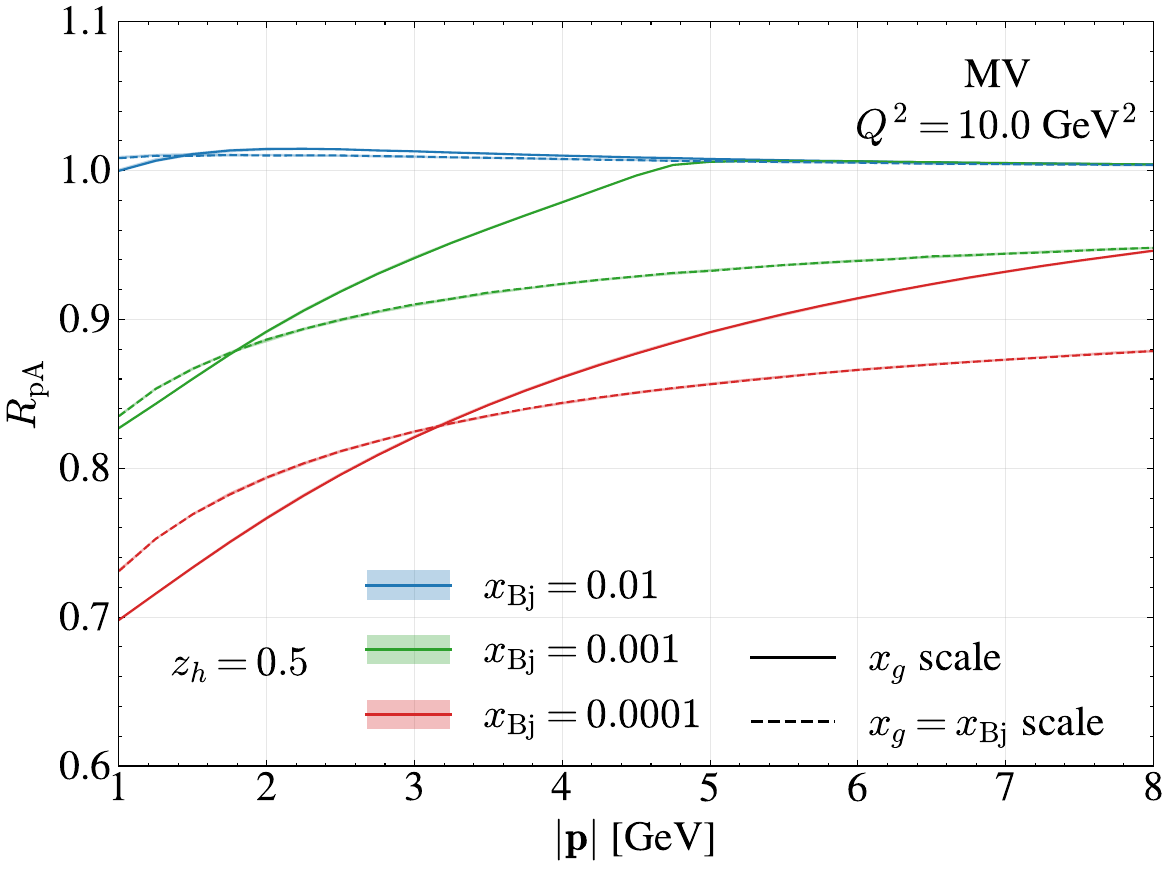}
\caption{MV}
\label{plot:Plot_SIDIS_nuclear_mod_EIC_R_vs_p_xg_Vs_xbj_MV}
\end{subfigure}
\hfill
\begin{subfigure}[t]{0.49\textwidth}
\centering
\includegraphics[width=\linewidth]{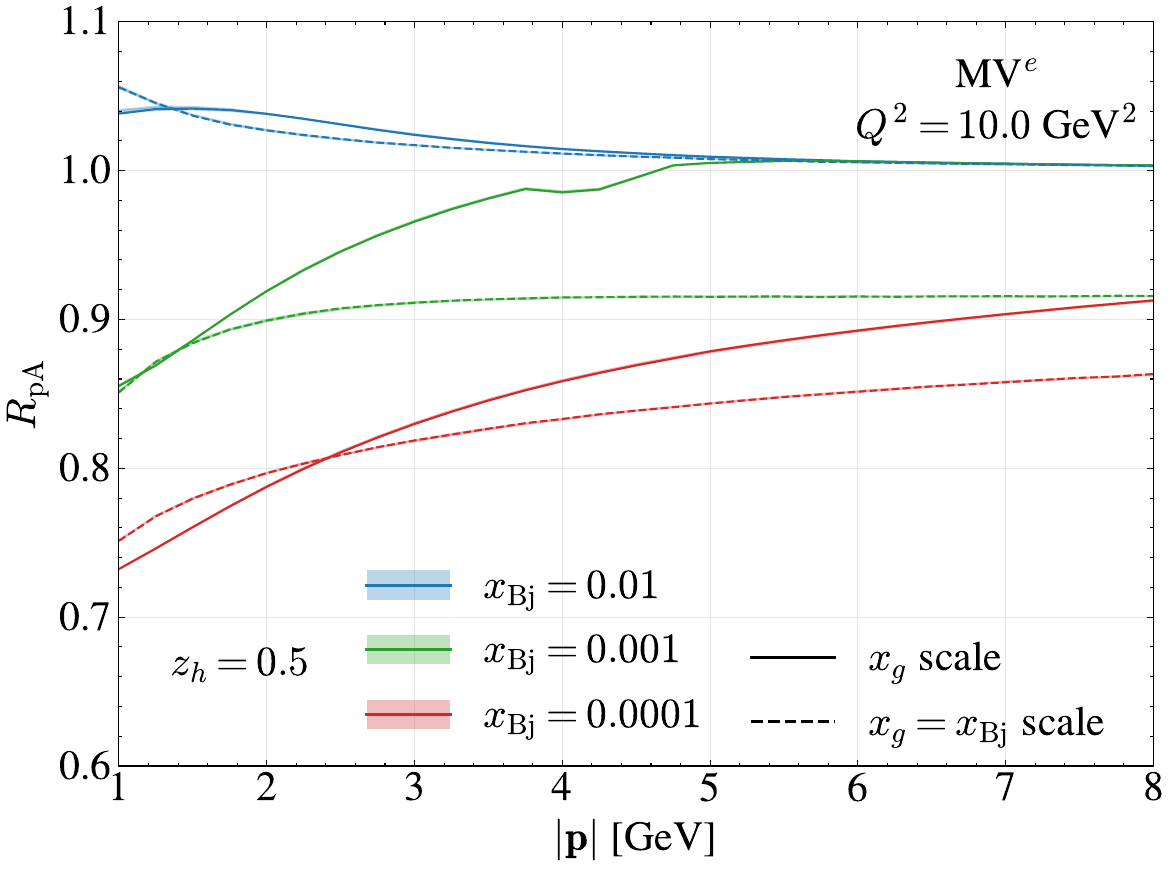}
\caption{MV$^e$}
\label{plot:Plot_SIDIS_nuclear_mod_EIC_R_vs_p_xg_Vs_xbj_MVe}
\end{subfigure}

\caption{Nuclear modification factor as a function of $|\pt|$ for selected values of $\xbj$ at fixed $Q^2 = 10$ GeV$^2$ and $z_h=0.5$. Results are shown for the MV and MV$^e$ model fits of dipole amplitudes. Solid lines correspond to the $x_g$ BK evolution scale defined in Eq.~\eqref{eq:xg} and dotted lines correspond to $x_g = \xbj$ BK evolution scale.
}
\label{fig:Plot_SIDIS_nuclear_mod_EIC_R_vs_p_xg_Vs_xbj}
\end{figure*}

We quantify this sensitivity in Fig.~\ref{fig:Plot_SIDIS_nuclear_mod_EIC_R_vs_p_xg_Vs_xbj} by calculating the nuclear 
 modification factors $R_\mathrm{pA}$ at a typical scale $Q^2=10\,\mathrm{GeV}^2$, using two different choices for the BK evolution range. 
 We compare our  default prescription, referred to as ``$x_g$ scale'', with a calculation in which the BK evolution is solved down to $\xbj$. Although the latter choice is consistent with that used in DIS structure function fits~\cite{Lappi:2013zma,Casuga:2023dcf}, but otherwise it is less well motivated physically than the estimate in Eq.~\eqref{eq:xg}, which is based on the typical invariant mass of the produced system.

By construction, the nuclear modification factors at $\xbj=0.01$ agree within numerical accuracy. 
At smaller values of $\xbj$, the choice of $x_g$ prescription produces a noticeable effect at large $|\pt|$, where both  
the parton transverse momentum and the invariant mass of the final state are large.
This sensitivity can be understood as follows. 
When using the $x_g$ scale of Eq.~\eqref{eq:xg}, the dipole amplitude is evaluated close to its initial condition at all $\xbj$ when the final state invariant mass is large. Consequently, $R_\mathrm{pA}$ approaches unity at high $|\pt|$. In contrast, evaluating the dipole amplitude at a $|\pt|$-independent scale determined by $\xbj$ leads to significant suppression across the full $|\pt|$ range.
The resulting  differences  can reach $\sim 0.1$ in $R_\mathrm{pA}$, exceeding the uncertainty associated with the BK initial condition parametrization quantified earlier in this work. At lower transverse momenta, $|\pt|\lesssim 3\,\mathrm{GeV}$, the dependence on the $x_g$ scale prescription is only moderate.

We expect the difference between these rapidity scale prescriptions to decrease once DIS fits, in which the BK evolution rapidity is determined by the invariant mass of the produced system, become available. Such fits can be performed by applying the developments of Ref.~\cite{Bertilsson:2026vtu}, combined with a realistic extrapolation to the $x_g>0.01$ regime.

\bibliographystyle{JHEP-2modlong.bst}
\bibliography{refs}

\end{document}